\documentclass[12pt,preprint]{aastex}

\usepackage{natbib}
\usepackage[latin1]{inputenc}
\usepackage{bm}
\usepackage{graphicx}
\usepackage{wasysym}
\usepackage{amssymb}

\newcommand{\pz}{\phantom{0}}
\newcommand{\pminus}{\phantom{-}}
\newcommand{\text}{\rm}

\shorttitle{A General Relativistic Model of Light Propagation}
\shortauthors{de Felice et al.}

\begin{document}

\title{A General Relativistic Model of Light Propagation \\
       in the Gravitational Field of the Solar System: the Static Case}
\author{de Felice F.\altaffilmark{1}, Crosta M.T. \altaffilmark{1,2} and 
        Vecchiato A.\altaffilmark{1,2}} 
\affil{Department of physics, University of Padova
       via Marzolo 8, 35131 Padova, Italy}
\and 
\author{Lattanzi M.G. and Bucciarelli B.}
\affil{Turin Astronomical Observatory
       strada Osservatorio 20, 10125 Pino Torinese (TO), Italy}
\altaffiltext{1}{Also at INFN - Sezione di Padova}
\altaffiltext{2}{Also at Turin Astronomical Observatory}


\begin{abstract}
We develop here a new approach for the relativistic modeling of the
  photons moving into a quasi-Minkowskian space-time, where the metric
  is generated by an arbitrary n-body distribution within an isolated
  Solar System.  Our model is built on the prescriptions of the theory
  of General Relativity and leaves the choices of the metric, as well
  as that of the motion of the observer, arbitrary.  Adopting a
  quasi-Minkowskian expression of the metric accurate to order
  $(v/c)^2$, a thorough numerical test campaign is conducted to verify
  correctness and reliability of the model equations.  The test
  results show that the model behaves according to predictions.
  Specifically, comparisons to true (simulated) data demonstrate that
  stellar distances are reconstructed up to the specified level of
  accuracy.  Although the $(v/c)^2$ approximation is not always
  sufficient for its application to future astrometric experiments,
  which require modeling to $(v/c)^3$, this work serves also as a
  natural test-ground for the higher order model, whose formulation is
  now close to completion, and will be presented in a forthcoming
  paper.
\end{abstract}
\keywords{relativity --- gravitation --- astrometry}

\section{\label{sec:intro}Introduction}
It is known that the light signal carries most of the physical
information on the celestial objects and the physical field it passed
through (see \citet{1999PhRvD..60l4002K,2002PhRvD..65f4025K} and
references therein). Part of this information, like e.g.\ the
distance to the source and its velocity, can be extracted by means of
(classical) astrometric techniques, but at high-level of accuracy, as
demonstrated by ESA space mission Hipparcos, relativistic effects
cannot be neglected, and this ``relativistic astrometry'' can be
considered part of fundamental physics (for example, astrometric
measurements are one of the natural choices to detect possible
deviations from General Relativity generated by a scalar field $\phi$
that couples with the metric tensor $g$ to generate gravity
\citep{2002PhRvL..89h1601D,2002PhRvD..66d6007D}, or they were used for
a possible attempt at measuring the speed of gravity
\citep{2001ApJ...556L...1K,2003ApJ...598..704F}.

The purpose of this paper is to construct a model of the celestial
sphere using the prescriptions of the General Theory of Relativity in
order to take into account the relativistic effects suffered by light
while propagating through the gravitational field of the Solar System.
The planned future astrometric experiments set the optimal target
accuracy for such models to the micro-arcsecond level.  To this level,
it was shown that the light propagation will be affected not only by
the mass of the Sun and of the other planets but also by their
gravitational quadrupole and their translational and rotational motion
\citep{1992AJ....104..897K,2002PhRvD..65f4025K}. Assuming that the
Solar System is isolated and it is the only source of gravity, then
the accuracy of $0.1~\mu$arcsec is achieved by considering terms of
the background metric up to the order of $(v/c)^3$; here $c$ is the
velocity of light in vacuum and $v$ is an average velocity for energy
balance.  In the Solar System this is of $\sim 10$ Km sec$^{-1}$.

In this sense, the next generation astrometric missions like GAIA, of
ESA \citep{2001AAp...369..339P,2003GSST.ASPcs...3P}, and SIM, of NASA
\citep{1998SPIE.3350..536S,2003SPIE.4852....1M}, offer a unique
opportunity to test our model.  On the other hand, the observational
error of these satellites will be pushed to the microarcsecond level
($\mu$arcsec) therefore, the need is to implement a model of the
celestial sphere and of the observables which is accurate to that
order. As a reference, we recall that Hipparcos (of ESA), the only
previous example of a global astrometric mission, had an accuracy of
1~milliarcsecond only.

The construction of a general relativistic many-body astrometric model
with the required accuracy is expected to be complicated not only by
the mathematical structure of the relativistic equations but also by
the numerical methods deployed to implement the model into software
code. Therefore it is crucial to have an efficient strategy for
testing the model. With this goal in mind, we had already developed a
relativistic astrometric model taking as background geometry the exact
(unperturbed) Schwarzschild solution
\citep{1998AAp...332.1133D,2001AAp...373..336D}. That model was used as
a basic touchstone for comparison in the construction of our many-body
model. Moreover, since a model limited to the $(v/c)^2$ order of
accuracy is significantly simpler to handle and test, we decided for a
full computer implementation of the $(v/c)^2$-model, with the
intention to use it as test-ground for the higher order extension,
which is our ultimate goal.

Finally, it is worth stressing that our aim was not to add to the 
theory of light propagation, which is well known, but to develop a 
relativistic astrometric model with a new theoretical approach that treats 
the light propagation in a curved space-time in such a way that all the 
possible relativistic perturbations coming from the gravity sources are 
naturally included in the light-path reconstruction.

In sections~\ref{sec:metric} and~\ref{sec:set-geometry}, we define the
background geometry in a way compatible with our basic physical
assumptions. In section~\ref{sec:light-traj} we discuss how to handle
light propagation through the Solar System and provide the
differential equations which enable us to reconstruct the light
trajectory from a distant star to the observer.  Then, in
sections~\ref{sec:incond} and~\ref{sec:emtime}, we define the {\it
  observables\/} as the measurements coming from a given satellite and
link them to the mathematical boundary conditions needed to integrate
the light trajectory.  Finally, in sections~\ref{sec:test}
and~\ref{sec:n-bod-test}, we present the test campaign used to
validate the model.

In what follows, Greek indices run from $0$ to $3$ and Latin indices
run from $1$ to $3$.

\section{\label{sec:metric}The space-time metric}
The space-time structure which underlies the development of
relativistic astrometry must mirror the physical and operational
reality experienced by the observer.

The basic step of this project is to identify the background geometry.
Our first assumption is that the Solar System is isolated; this means
that there are no perturbing bodies intervening between the emitting
stars and the Solar System boundaries. It is clear that this
assumption may not be fully justified since light rays from distant
stars may suffer deflections due to \textit{microlensing} induced by
intervening bodies (e.g., as far as the mission GAIA is concerned, we
expect nearly 1000 microlensing events along the Galactic disk during
5 years \citep{2000ApJ...534..213D}); indeed they could generate
systematic errors in the data reduction of the celestial sphere,
however, their number is very small as compared to the number of
observations ($\ge 10^9$).  Furthermore, we know that most of the
stars are in a binary system and have proper motion which generates an
unsteady gravitational field causing a rotation of the celestial
reference frame. However, the rotation coefficients have a temporal
variation which amounts to 1 $\mu$arcsec every 20 years, a period much
longer than the lifetime of all the future space astrometric missions
\citep{1998MNRAS.300..287S}.  So we judge that the above cases do not
affect the validity of the hypothesis that the Solar System is
isolated.

A second assumption is that the Solar System generates a weak
gravitational field.  We can then adopt a quasi-Minkowskian coordinate
system so that the space-time metric can be written as a perturbation
of the Minkowski metric $\eta_{\alpha\beta}$, namely:
\begin{equation}
g_{\alpha \beta}= \eta_{\alpha \beta} + h_{\alpha \beta}+O(h^2)
\label{eq:met}
\end{equation}
where the ${h_{\alpha\beta}}'s$ describe effects generated by the
bodies of the Solar System and are {\it small\/} in the sense that
$|h_{\alpha\beta}|\ll 1$, their spatial variations are of the order of
$|h_{\alpha\beta}|$ while their time variations are at most of the
order $(v/c)|h_{\alpha\beta}|$. Clearly the metric form (\ref{eq:met})
is preserved under gauge transformations of the order of $h$.  The
order of magnitude of the correction terms entering the
${h_{\alpha\beta}}'s$, is expressed in terms of powers of $(v/c)$.
The ${h_{\alpha\beta}}'s$ are at least of the order of $(v/c)^2$
(Newtonian terms) hence any higher level of accuracy within the model,
is fixed by the power of $(v/c)$ larger than two that one likes to
include in the analysis.

Having the above considerations in mind, we choose a solution of
Einstein's equations of the type:
\begin{equation}
g_{\alpha \beta}= \eta_{\alpha \beta}+ \sum_a h^{(a)}_{\alpha
\beta}+O(h^2) \label{eq:metsum}
\end{equation}
where now the sum is extended to the bodies of the Solar System.  In
this approximation, the metric tensor (\ref{eq:metsum}) has in general
a non-vanishing term $g_{0i} = O[(v/c)^3]$ and the non-linearity of
the gravitational field is confined to terms $O[(v/c)^4]$ in $h_{00}$
and $h_{ij}$.

Moreover, in August 2000 the General Assembly of the IAU stated that a
solution like (\ref{eq:metsum}) has to be adopted to define the
reference frames and time scales in the Solar System
\citep{1991PhRvD..43.3273D,1989NCimB.103...63B}.  At the first
Post-Newtonian level of approximation, the metric tensor nearby any
planet of the Solar System takes the form:
\begin{eqnarray}
g_{00}&=& -1 + \frac{2W}{c^2} -\frac{2W^2}{c^4}+ O(c^{-5})\label{eq:zelig1}\\
g_{0 i}&=& -\frac{4W_i}{c^3} + O(c^{-5}) \label{eq:zelig2}\\
g_{ij}&=&\delta_{ij} \Big( 1 + \frac{2W}{c^2} \Big)+ O(c^{-4})\label{eq:zelig3} 
\end{eqnarray}
with $i, j$ = 1, 2, 3.

In the equations above, $W$ represents a generalization of the
Newtonian potential and $W_i$ is a vector potential describing the
dynamical contribution to the background geometry by the relative
motion of the gravitating sources as well as by the peculiarities of
their extended structures.  The same form (\ref{eq:zelig1}) to
(\ref{eq:zelig3}) is adopted to describe the metric tensor generated
by the whole Solar System referring so forth to the Barycentric
Celestial Reference System. Assuming the space-time to be
asymptotically flat, the potentials $W$ and $W_i$ are a sum of
integrals containing terms of gravitational mass and mass-current
density.  Such integrals are taken over the support of each body of
the Solar System. It should be stressed here that the linearized form
of the metric, in the weak field and slow motion ($v\ll c$)
approximation, is the current support for calculating the ephemerides
at the Jet Propulsion Laboratory \citep{2000tmcs.conf..283S},
California Institute of Technology.

A light ray, on its way to the observer from a distant star, first
feels the gravitational field of the Solar System as a point like mass
centered in its barycenter, then, as it gets closer, it feels the
gravitational perturbations of the individual bodies of the System.
To the order of $(v/c)^2$, the light ray will feel the contributions
from the individual mass structures while the effects arising from
their relative motion and spin would enter terms of higher order.
Since a time derivative adds a factor of the order of $(v/c)$, time
derivatives of the metric coefficients generate terms at least of the
order of $(v/c)^3$.  Our third assumption then is that light rays will
not feel perturbations of the order O$[(v/c)^3]$ so, within each
integration of the light trajectories, the Solar System is
approximated in our model to a static, non-rotating and non-expanding
gravitating system. A natural consequence of this assumption is that,
along with time variations of the metric, we shall also neglect mixed
metric coefficients of type $h_{0i}$.  Furthermore, in calculating the
background metric coefficients which affects the light signal at each
spatial point of its trajectory, we need not to consider the spatial
location of the individual gravitational source at the corresponding
retarded time. This does not mean that the positions of the perturbing
bodies are considered fixed during the whole integration, but
simply that the ephemeris used do not include the retarded time, and
that we neglect, to $(v/c)^2$, the effects due to their
velocities.

\section{\label{sec:set-geometry}Setting the geometry of the celestial sphere}
Let ($\xi^i,\xi^0\equiv \tau$) be a quasi-Minkowskian coordinate
system with respect to which the space-time metric takes form
(\ref{eq:met}); as stated, the choice to consider only terms up to
$(v/c)^2$ amounts to assume that space-time is static. Hence there
exists a time-like Killing vector field $\eta$, say, along which the
physical properties of the space-time do not change and therefore
uniquely identifies a time direction. Let the coordinate time $\tau$
be a parameter along $\bm{\eta}$ so that
$\eta^\alpha=\delta^\alpha_0$.  From the Killing equation
$\eta_{(\alpha;\beta)}=0$, where semicolon means covariant derivative
with respect to the given metric and round brackets mean
symmetrization, one easily deduces that the congruence $C_{\bm{\eta}}$
of Killing lines is, to the required order, vorticity-free. In fact,
denoting as
$P(\eta)^\alpha{}_\beta=\delta^\alpha_\beta+\eta^\alpha\eta_\beta$ the
tensor operator which projects orthogonally to $\bm{\eta}$, one finds:
\begin{equation}
\label{eq:vort}
\omega_{\alpha\beta}=P(\eta)^\rho{}_{[\alpha}P(\eta)^\sigma{}_{\beta]}\eta_{\rho;\sigma}=O[(v/c)^3]
\end{equation}
where $\omega_{\alpha\beta}$ is the vorticity tensor and square
brackets mean anti-symmetrization. A space-time which admits a
vorticity-free congruence of lines can be \textit{foliated}, namely it
admits a family of three-dimensional space-like hypersurfaces
$S(\xi^i,\tau)$ which are everywhere orthogonal to $C_{\bm{\eta}}$. It
is always possible to choose a coordinate system such that the
surfaces $S(\xi^i,\tau)$ have equation $\tau=$constant; in this case
the spatial coordinates can be fixed within each slice up to spatial
transformations only. We shall term these surfaces $S(\tau)$.

\begin{figure}[htp]
\plotone{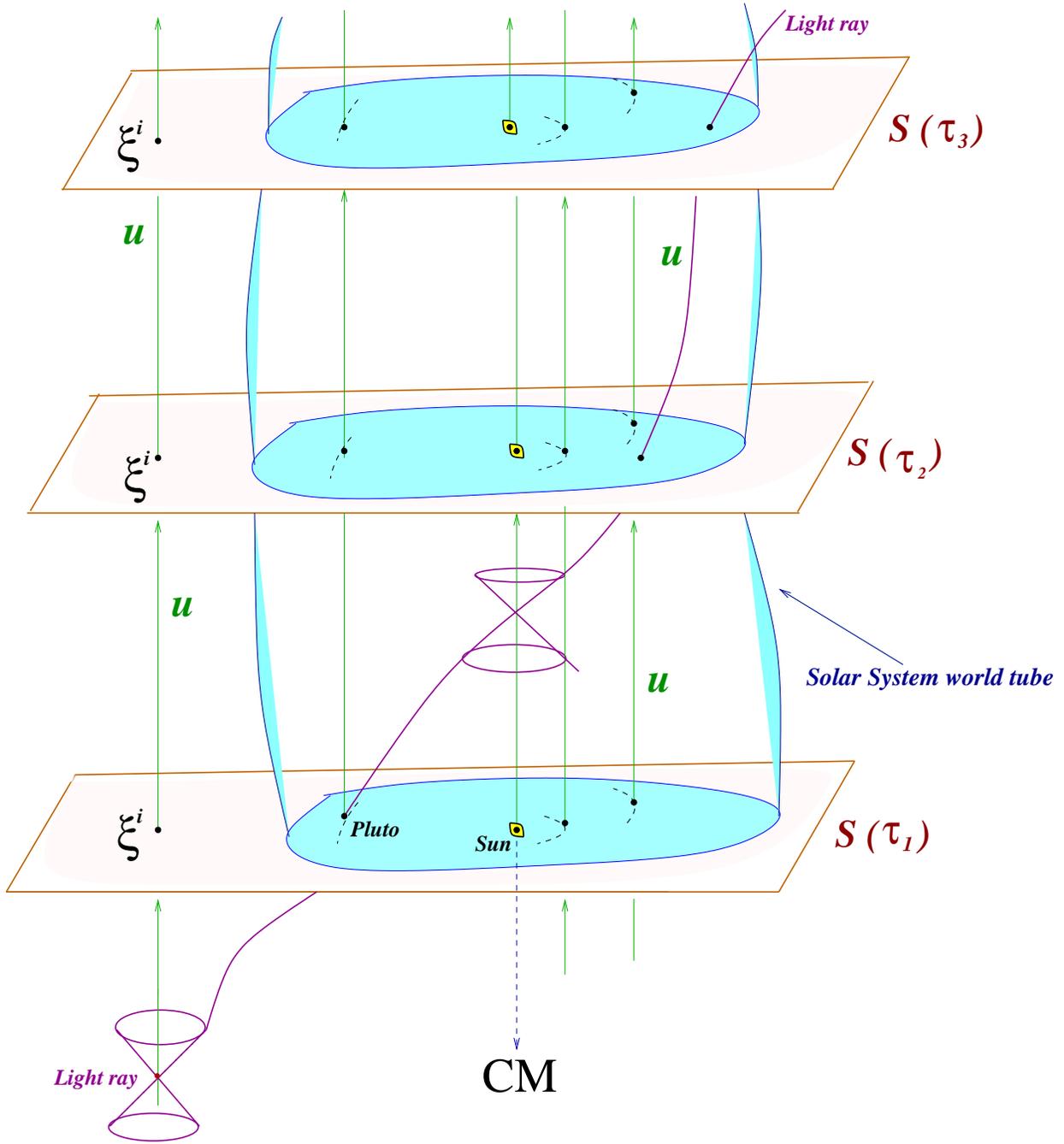}
\caption{\label{fig:sswt}In our model each light ray behaves as 
  if the bodies did not move with respect to the center of mass CM.}
\end{figure}

Let us now consider the unit vector field $\bm{u}$ which is everywhere
proportional to $\bm{\eta}$, namely $u^\alpha=
{\rm e}^\psi\delta^\alpha_0$ where $\text{e}^\psi=(-g_{00})^{-1/2}$ is
the normalization factor which assures that $u_\alpha u^\alpha=-1$;
the associated one-form has components $u_\alpha=-{\rm
  e}^{-\psi}\frac{\partial\tau}{\partial\xi^\alpha}$ so it is
proportional to the gradient of $S(\tau)$, as expected.  Through each
point of a slice $S(\tau)$, for any $\tau$, there goes a time-like
curve, orthogonal to $S(\tau)$ and having as tangent a vector of the
vector field ${\bf u}$; the totality of these curves through all the
points of $S(\tau)$ form a non-intersecting family of curves, or a
congruence $C_{\bf u}$, which identifies a physical observer.  The
property of this observer is to be static with respect to the selected
coordinate representation, that is the spatial coordinates do not vary
along its world-lines.  The parameter $\sigma$ on the congruence, such
that $u^\alpha=d\xi^\alpha/d\sigma$, is the proper-time of the
observer $\bm{u}$. We then require that the geometry that each photon
{\it feels\/} before reaching the target is described by metric
(\ref{eq:met}); moreover we also require that within each photon
travel time the world-lines of the bodies of the Solar System belong
to the congruence $C_{\bf u}$ and in particular the barycenter of the
Solar System is fixed at the origin of the spatial coordinates on each
slice, (\figurename~\ref{fig:sswt}).  With this choice, the observer
${\bf u}$ will be termed {\it locally barycentric}. This observer is
an essential prerequisite of our relativistic astrometric model
because at any space-time point and apart from a position-dependent
rescaling of its time rate, it plays the role of the barycentric
observer which is located at the origin of the spatial coordinates
fixed, as said, at the barycenter of the Solar System. Before
concluding this section, let us recall here that the constraints on
the metric, namely $h_{0i}=0$ and $C_{\bm{\eta}}$ being Killing and
vorticity-free, are gauge invariant only up to the order of $(v/c)^2$.

\clearpage
\clearpage


\section{\label{sec:light-traj}The light trajectories}
\begin{figure}[htp]
\plotone{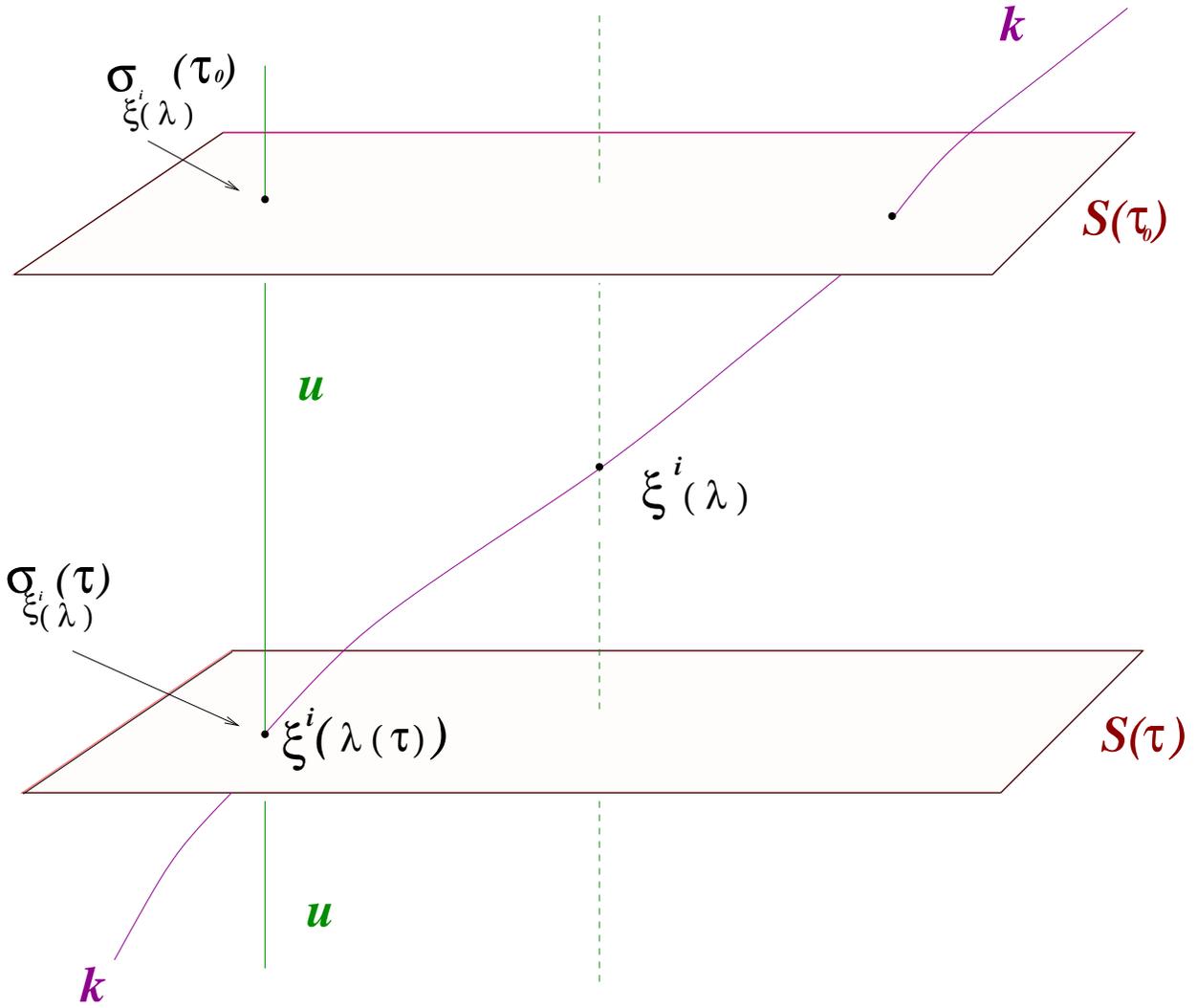}
\caption{\label{fig:sigmalambda}The point $\xi^i(\lambda(\tau))$ belongs to
  the null geodesic but also to the unique normal to the slice $S(\tau)$
  crossing it with a value of the parameter
  $\sigma_{\xi^i(\lambda)}(\tau)$.}
\end{figure}

A photon traveling from a distant star to the observer within the
Solar System, would see the space-time as a time development of slices
of constant $\tau$. As stated, we shall treat each light trajectory
assuming that the bodies of the Solar System were fixed at the spatial
position they had at the time of observation, say.  Evidently, any
subsequent light ray will be considered updating the positions of the
bodies of the Solar System according to their actual motion.  Let us
then prepare the way to a suitable treatment of a light ray.  Let
\begin{equation}
P(u)_\alpha{}^\beta = \delta_\alpha{}^\beta+u_\alpha u^\beta 
\label{eq:proj2}
\end{equation}
be the operator which projects orthogonally to $\bm{u}$. Because of
the unitary condition, the parameter $\sigma$ on the trajectories of
$\bm{u}$ is not constant on the slices $S(\tau)$ but varies
differentially with the position as $\sigma=\sigma(\xi^i,\tau)$. Since
the spatial coordinates $\xi^i$ are constant along the unique normal
going through the point with those coordinates, the parameter $\sigma$
along it will be function of $\tau$ only; i.e.
$\sigma=\sigma_{\xi^i}(\tau)$.  Let us now consider a null geodesic
$\mit\Upsilon$ with tangent vector field $k^\alpha\equiv
d\xi^\alpha/d\lambda$ which satisfies the following equations:
\begin{eqnarray}
\label{eq:geodes} 
k^\alpha k_\alpha&=&0\\ 
\frac{dk^\alpha}{d\lambda}+{\Gamma}^{\alpha}_{\rho\sigma} k^\rho k^\sigma&=&0;\nonumber
\end{eqnarray}
the latter express respectively the null and the geodesic conditions;
here $\lambda$ is a real parameter on $\mit\Upsilon$ and
$\Gamma^{\alpha}_{\rho\sigma}$ are the connection coefficients of the
given metric (\figurename~\ref{fig:sigmalambda}).  Assume that the
trajectory starts at a point $P_\ast$ on a slice $S(\tau_*)$ (say) and
with spatial coordinates $\xi_*^i$.  The light trajectory will {\it
  end\/} at the observation place on a slice $S(\tau_0)$ and at a point
with spatial coordinates $\xi_{(0)}^i$. We remember that the origin of
the coordinate system is meant to be the barycenter of the Solar
System.  The purpose of this work is to determine $\xi_*^i$, namely
the coordinates of the star, from a prescribed set of observables.
\clearpage
\clearpage

Since our approximation permits a global foliation of the space-time,
we find it more convenient for the determination and physical
interpretation of the gravitational effects on light propagation, to
consider the {\it spatial projection\/} of the light ray $\mit\Upsilon$
on the slice $S(\tau_0)$; this projection is a curve made of the
points of $S(\tau_0)$ which one gets to by moving along the unique
normal through the point of intersection of the light ray with the
slice $S(\tau)$, for any $\tau$. This curve, that we denote as
$\bar{\mit\Upsilon}$, is smooth and has a tangent vector field
${\bf\ell}$ whose coordinate components are equal to those of the
projection of the tangent ${\bf k}$ to the null ray into the rest
frame of the local barycentric observer at each of its points, namely
\begin{equation}
\ell^\alpha=P(u)^\alpha{}_\beta k^\beta
\end{equation}
(see \appendixname~\ref{app:mapping} for a mathematical description of
this mapping procedure). The vector ${\bf\ell}$ physically identifies
the local line of sight of the local barycentric observer at each
space-time point of the light ray trajectory. Indeed, the knowledge of
${\bf \ell}$ is essential to reconstruct the whole story of the light
ray. The curve $\bar{\mit\Upsilon}$ will be naturally parametrized by
$\lambda$.

From $u^\alpha u_\alpha=-1$, it follows that:
\begin{equation}
\ell^\alpha=k^\alpha+u^\alpha (u_\beta k^\beta)\,. \label{eq:ellsplit}
\end{equation}
Clearly it is $\ell^0=0$ showing that $\ell^\alpha$ lies everywhere on
the slice $S(\tau_0)$ as expected.  Since each point of
$\bar{\mit\Upsilon}$ is the image under the above projection operation
of a point of $\mit\Upsilon$ at time $\tau$, it is more convenient to
label the points of $\bar{\mit\Upsilon}$ with the value of the
parameter $\sigma_{\xi^i(\lambda)}(\tau)$ which, as we have already
said, uniquely identifies that point on the normal to the slice
$S(\tau)$ which contains it.  Hence, being
\begin{equation}
d\sigma=-(u_\alpha k^\alpha) d\lambda\,,
\label{eq:newpar}
\end{equation}
we define the new tangent vector field:
\begin{equation}
\bar\ell^\alpha\equiv \frac{d\xi^\alpha}{d\sigma_{\xi^i(\lambda)}}= -\frac{\ell^\alpha}{\Big(u^\beta k_\beta\Big)}.\label{eq:ellbar}
\end{equation}
In the same way we denote
\begin{equation}
\label{eq:kappabar}
\bar k^\alpha\equiv -\frac{k^\alpha}{(u_\beta k^\beta)}
\end{equation}
so that
\begin{equation}
\bar k^\alpha=\bar\ell^\alpha+u^\alpha \label{eq:kapsplit}
\end{equation}
which implies:
\begin{equation}
\bar\ell^{\alpha } \bar\ell_{\alpha}=1. \label{eq:ellun}
\end{equation}
In what follows we shall denote $\sigma_{\xi^i(\lambda)}$ as $\sigma$.
We can now write the differential equation which is satisfied by the
vector field $\bar{\bm{\ell}}$. From (\ref{eq:newpar}) and
(\ref{eq:kappabar}), the second of equations (\ref{eq:geodes}) writes:
\begin{eqnarray}
\frac{d \bar{\ell}^{\alpha}}{d\sigma}&+& \frac{d u^{\alpha}}{d\sigma}- 
(\bar{\ell}^{\alpha}+u^{\alpha})(\bar{\ell}^{\beta}\dot{u}_{\beta}+
\bar{\ell}^{\beta}\bar{\ell}^{\tau}\nabla_{\tau}u_{\beta})\nonumber\\
&+&\Gamma^{\alpha}_{\beta \gamma}(\bar{\ell}^{\beta}+
u^{\beta})(\bar{\ell}^{\gamma}+u^{\gamma})=0\,. 
 \label{eq:geoC}
\end{eqnarray}
 
In this equation the quantity
$\bar{\ell}^{\beta}\bar{\ell}^{\tau}\nabla_{\tau}u_{\beta}$ can be
written explicitly in terms of the expansion $\Theta_{\rho \sigma}$ of
$C_{\bm{u}}$ \citep{1990recm.book.....D}, as:
\begin{equation}
\bar{\ell}^{\beta}\bar{\ell}^{\tau}\nabla_{\tau}u_{\beta}=
\Theta_{\rho \sigma} \bar{\ell}^{\rho}\bar{\ell}^{\sigma}
\end{equation}
where
$\Theta_{\rho\sigma}=P(u)^\alpha{}\rho P(u)^\beta{}_\sigma\nabla_{(\alpha}u_{\beta)}$
Since the only non vanishing components of the expansion are
$\Theta_{ij}=(1/2)\partial_0 h_{ij}$, the expansion vanishes
identically as consequence of the assumption to neglect time
variations of the metric.  From this condition and imposing $h_{0i}=0$
equation (\ref{eq:geoC}) becomes to the required order and after some
algebra:

\begin{eqnarray}
\frac{d \bar{\ell}^{\alpha}}{d\sigma}&+&\frac{1}{2}
\left(\bar{\ell}^{i}\partial_{i}h_{00}\right)
\delta^{\alpha}_{0}+\frac{1}{2}\left(\bar\ell^i\partial_i h_{00}\right)
\left(\bar\ell^\alpha+\delta^\alpha_0\right)\nonumber\\
&+&\eta^{\alpha\sigma}\left(\partial_ih_{\sigma j}-\frac{1}{2}
\partial_\sigma h_{ij}\right)\bar\ell^i\bar\ell^j+\eta^{\alpha\sigma}
\left(\partial_ih_{0\sigma}\right)\bar\ell^i\nonumber\\
&-&\frac{1}{2}\eta^{\alpha\sigma}\partial_\sigma h_{00}=0
 \label{eq:geowexp}
\end{eqnarray}

If $\alpha=0$, equation (\ref{eq:geowexp}) leads to
$\frac{d\bar\ell^0}{d\sigma}=0$ assuring that condition $\bar\ell^0=0$
holds true all along the curve $\bar{\mit\Upsilon}$; if $\alpha=k$
equation (\ref{eq:geowexp}) gives the set of differential equations
that we need to integrate to identify the coordinate position of the
star:

\begin{eqnarray}
\frac{d\bar{\ell}^{k}}{d\sigma}&+&\bar{\ell}^{k} \left(\frac{1}{2}\bar{\ell}^{i}\partial_{i}h_{00}\right)+\delta^{k s}\left(\partial_{i} h_{s j}-\frac{1}{2}\partial_{s} h_{ij}\right)\bar{\ell}^{i}\bar{\ell}^{j} \nonumber \\
&-&\frac{1}{2}\delta^{ks}\partial_{s} h_{00}=0,\nonumber\\
\bar\ell^k&=&\frac{d\xi^k}{d\sigma}. \label{eq:geodint}
\end{eqnarray}
Here we remind that Latin indexes take values $ 1, 2, 3$.

\section{\label{sec:incond}Observables and boundary conditions}
\begin{figure}
\plotone{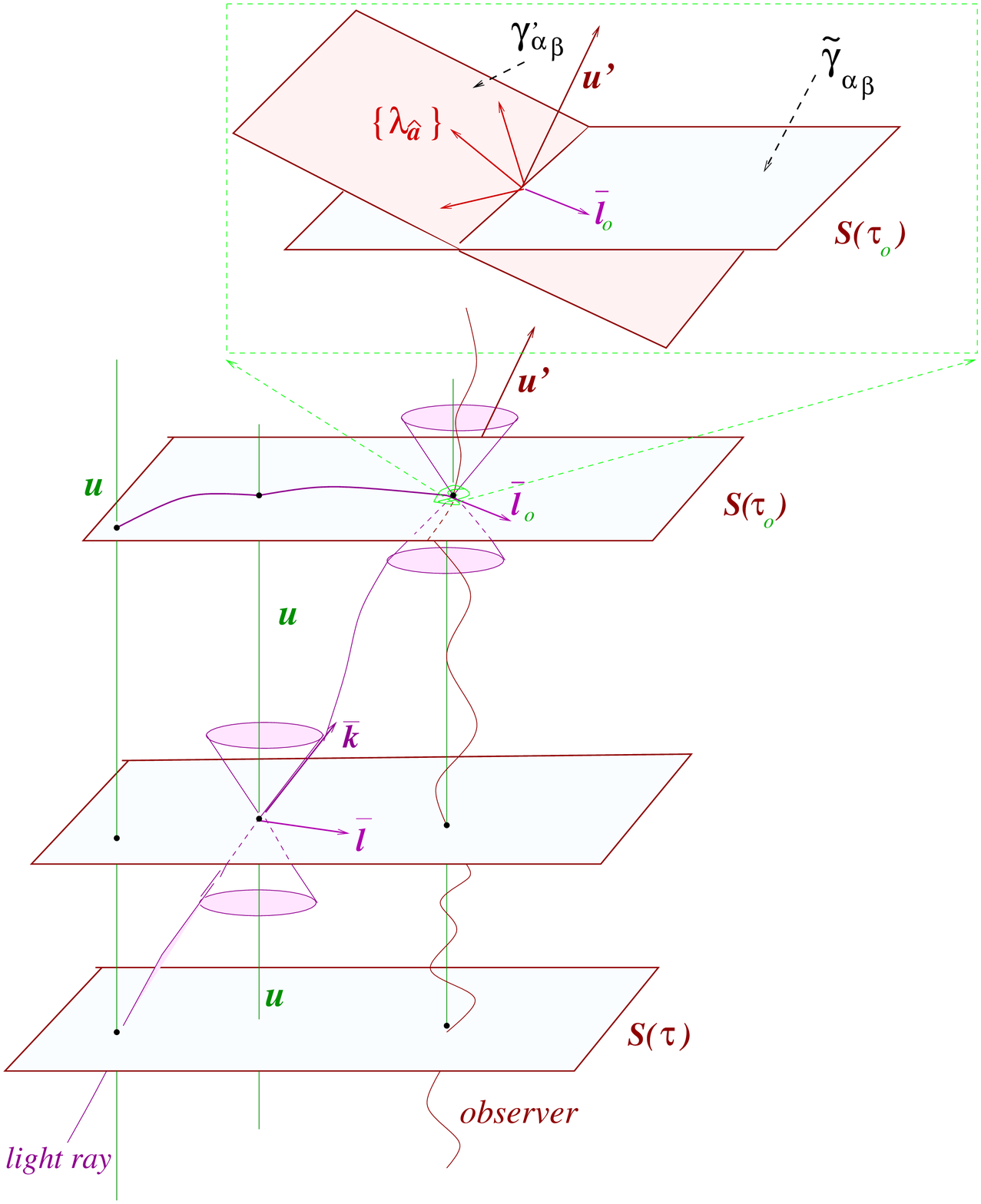}
\caption{\label{fig:tetrade}Fixing the boundary condition at 
  the time of the observation $\tau_0$, where $u'$ is the 
  tangent vector to the world line of the observer.}
\end{figure}

Our aim is to determine the coordinate positions of a star, from a
prescribed set of observables. As first step we shall express the
observables in a way which reflects a convenient set up for the
observer.  The latter carries a frame, namely a clock which measures
its proper-time and a rest-space, which are different from the
barycentric proper-time and rest-space, respectively given by the
parameter $\sigma(\tau)$ along the congruence $C_{\bm u}$ and the
space-like slices $S(\tau)$. To determine the boundary conditions
needed to integrate equation (\ref{eq:geodint}), we need both frames;
the observer frame allows us to define the measurements and the
barycentric one to express all the coordinate tensorial components.
Each measurement is made at a coordinate time $\tau_0$ when the
observer was at a spatial position with respect to the barycenter
given by the coordinates $\xi^i_{(0)}$.  Hence we consider as
\textit{observables} the angles between the direction of the incoming
photon and the three spatial directions of a frame adapted to the
observer. These three angles provide the required boundary values for
$\bar\ell^i$.  Let $\bm{u'}$ be the vector field tangent to the
observer's world-line and let $\{\bm{\lambda_{\hat{a}}}\}$ (where
$\hat{a}= 1, 2, 3$) be a space-like triad carried by the observer.
The angles $\psi_{(\lambda_{\hat{a}},\bar\ell)}$ that the incoming
light ray forms with each of the triad direction, is given by
\citep{1990recm.book.....D}:

\begin{equation}
\label{eq:cos}
\cos\psi_{(\lambda_{\hat{a}},\bar\ell)}  \equiv \textbf{e}\bm{'_{\hat{a}}}= \frac{\gamma'_{\alpha \beta} \bar k^{\alpha}\lambda^{\beta}_{\hat{a}}}{(\gamma'_{\alpha \beta} \bar k^{\alpha}\bar k^{\beta})^{1/2} (\gamma'_{\alpha \beta}  \lambda^{\alpha}_{\hat{a}}\lambda^{\beta}_{\hat{a}})^{1/2}}
\end{equation}
where no sum is meant over $\hat a$ and
${\gamma'_\alpha}^{\beta}=\delta_\alpha^\beta+u^{'\beta}u'_\alpha$ is
the operator which projects into the observer's rest-frame
(\figurename~\ref{fig:tetrade}).
\clearpage
\clearpage

From (\ref{eq:kapsplit}), the above equation can be written more
conveniently as
\begin{equation}
\textbf{e}\bm{'_{\hat{a}}}=
\frac{\bar{\ell}^i_{0}\lambda_{i\hat{a}}+ 
u^\beta\lambda_{\beta\hat{a}}}{{u'}_j \bar{\ell}_{0}^j + u^{'\rho} u_\rho}
\label{eq:Chata}
\end{equation}
where $\bar\ell_{0}\equiv\bar\ell(\tau_0)$ and recalling that
$\gamma'_{\alpha\beta}\lambda^\alpha_{\hat a}\lambda^\beta_{\hat
  a}=1$; this is a matrix equation where the unknowns are the photon's
spatial directions $\bar{l}^{k}_{0}$ at the time of observation; they
can be singled out as
\begin{equation}
\bar{\ell}^i_{0}[u'_i\textbf{e}\bm{'_{\hat{a}}}-\lambda_{i\hat{a}}]=u^\alpha\lambda_{\hat{a}\alpha}-\textbf{e}\bm{'_{\hat{a}}}(u^{'\alpha}u_\alpha). \label{eq:3conl}
\end{equation}

The direction of the light ray, as it is seen from the point of
observation, depends on the motion of the latter relative to the
center of mass.  This dependence gives rise to the stellar
\textit{aberration}.  Did the observer not move with respect to the
spatial coordinate grid, namely if $u^{'\alpha}=u^\alpha$, then
(\ref{eq:3conl}) would become $\bar\ell^i_{0}\lambda_{i\hat
  a}=\textbf{e}\bm{_{\hat a}}$ as expected.  Equations
(\ref{eq:geodint}) and the three conditions (\ref{eq:3conl}) together
with equations (\ref{eq:ellbar}) and the coordinate positions
$\xi^i_{(o)}=\xi^{i}(\tau_{o})$ of the observer at the time of
observation, form a closed system of equations whose solutions are the
coordinate positions of the star provided one is able to identify the
value of the parameter $\sigma$ at the emission. The solution of
(\ref{eq:geodint}), in fact, is of the type
$\xi^i(\sigma,\,\bar\ell_{0}^i,\,\xi_{(o)}^i)$ hence
$\sigma_*\equiv\sigma(\tau_*)$ which marks the photon emission, is an
implicit unknown. The way how to determine $\sigma_*$ will be
discussed in the following section.

In order to find the boundary conditions $\bar\ell^i_{0}$ for
equations (\ref{eq:geodint}), we have first to construct a {\it
  tetrad\/} frame adapted to the observer. The tetrad frame\-
\{$\bm{\lambda_{\hat{\alpha}}}$\} forms a system of local Cartesian
axes, which equips the observer with an \textit{instantaneous inertial
  frame}; in fact it is:
\begin{equation}
\left( \bm{\lambda_{\hat{\alpha}}} \mid  \bm{\lambda_{\hat{\beta}}} \right)= \eta_{\hat{\alpha} \hat{\beta}}. \label{eq:inercon}
\end{equation}
Let us relabel for convenience the coordinates $\xi^\alpha$ as
$(\tau,\, x,\, y,\, z)$; each tetrad vector
$\bm{\lambda_{\hat\alpha}}$ can then be expressed in terms of
coordinate components with respect to the coordinate basis,
as\footnote{The letter $s$ is for ``satellite-observer''}:

\begin{eqnarray}
\bm{u'}\equiv\bm{\lambda_{\hat{0}}}&=&\text{e}^{\psi '} \left( T_s \bm{\partial_0} + X_s \bm{\partial_x} + Y_s \bm{\partial_y} + Z_s \bm{\partial_z} \right) \label{eq:lamb0}\\
&{}&\nonumber\\
\bm{\lambda_{\hat{a}}}&=& T_a \bm{\partial_0} + X_a \bm{\partial_x} + Y_a \bm{\partial_y} + Z_a \bm{\partial_z}  \label{eq:lamba}
\end{eqnarray}
where $\text{e}^{\psi'}$ is the normalization function which makes
$\bm{u'}$ unitary and $a=1,\,2,\,3$.  Conditions (\ref{eq:inercon})
must be simultaneously satisfied hence after some algebra we obtain:
\begin{equation}
\text{e}^{\psi'} = \left[-\left( T_{s}^{2} g_{00} + X_{s}^{2} g_{xx} +
Y_{s}^{2} g_{yy} +Z_{s}^{2} g_{zz}\right)\right]^{-1/2}. \label{eq:fatns}
\end{equation}
and
\begin{eqnarray}
   T_s T_{a} g_{00}+X_{s} X_{a} g_{xx}+Y_{s} Y_{a} g_{yy}+Z_{s} Z_{a} g_{zz}=0 \label{eq:tricona} \\
   T_{a} T_{b} g_{00}+X_{a} X_{b} g_{xx}+Y_{a} Y_{b} g_{yy}+Z_{a} Z_{b} g_{zz}=0 \label{eq:triconb} \\
   T_{a}^{2} g_{00}+X_{a}^{2} g_{xx}+Y_{a}^{2} g_{yy} +Z_{a}^{2} g_{zz}=1 \label{eq:triconc}
\end{eqnarray}
where $ a\neq b $. A general solution of (\ref{eq:tricona}--\ref{eq:triconc}) is given by:
\begin{eqnarray}
\bm{\lambda_{\hat{0}}}&=& \text{e}^{\psi '} \left(\bm{\partial}_0+X_s\bm{\partial_x}+Y_s\bm{\partial_y}+Z_s\bm{\partial_z}\right) \label{eq:lamb0g} \\
\bm{\lambda_{\hat{1}}}&=& X_1\bm{\partial_x}+Y_1\bm{\partial_y} \label{eq:lamb1g} \\
\bm{\lambda_{\hat{2}}}&=& T_2\bm{\partial_0}+X_2\bm{\partial_x}+Y_2\bm{\partial_y}+Z_2\bm{\partial_z} \label{eq:lamb2g} \\
\bm{\lambda_{\hat{3}}}&=& T_3\bm{\partial_0}+Z_3\bm{\partial_z} \label{eq:lamb3g}
\end{eqnarray}
where the components are explicitly given in
\appendixname~\ref{app:tetr-comp}.  This solution describes the
instantaneous inertial frame of an observer endowed with a general
motion in a gravitational field described by metric (\ref{eq:metsum}).

It is crucial that our model is consistent, under the same conditions,
with the one discussed in \citet{1998AAp...332.1133D} and
\citet{2001AAp...373..336D} where the Sun was the only source of
gravity.  In order to make comparison easier let us assume that the
observer moves on a circular orbit around the barycenter of the
Sun-Earth system and has, at the time of observation, spatial
coordinates $x_0, y_0,$ and $ z_0=0$.  The 4-velocity of the observer
now reads:
\begin{equation}
\bm{u}^{'\alpha}=\text{e}^{\psi'}\left(\bm{\partial_0}-\omega 
(y_0-y_\odot)\bm{\partial_x}+\omega(x_0-x_\odot)\bm{\partial_y}\right) 
\label{eq:smcu}
\end{equation}
where $\omega$ is the coordinate Keplerian angular velocity of
revolution and ($x_\odot,\,y_\odot$) are the coordinates of the
barycenter of the Sun-Earth with respect to the barycenter of the
Solar System.  In this case, the tetrad frame becomes:
\begin{eqnarray}
   \bm{\lambda_{\hat{0}}}\!\! &=&\!\! \text{e}^{\psi'}\left(\bm{\partial_0}-\omega(y_0-y_\odot)\bm{\partial_x}+\omega(x_0-x_\odot)\bm{\partial_y}\right) \label{eq:lamb0c} \\
   \bm{\lambda_{\hat{1}}}\!\! &=&\!\! X_1\bm{\partial_x}+Y_1\bm{\partial_y} \label{eq:lamb1c} \\
   \bm{\lambda_{\hat{2}}}\!\! &=&\!\! T_2\bm{\partial_0}+X_2\bm{\partial_x}+Y_2\bm{\partial_{y}} \label{eq:lamb2c} \\
   \bm{\lambda_{\hat{3}}}\!\! &=&\!\! Z_3\bm{\partial_z} \label{eq:lamb3c}
\end{eqnarray}
where:
\begin{equation}
\text{e}^{\psi'}\!\!=\!\!\left[-\!\left(g_{00}\!+\!\omega^2(y_0\!-\!y_\odot)^{2}g_{xx}\!+\!\omega^2 (x_0\!-\!x_\odot)^{2} g_{yy}\right)\right]^{-1/2}, \label{eq:fatnsc}
\end{equation}
\begin{eqnarray}
   X_1 &=& -\frac{\sqrt{g_{yy}} \, (x_0-x_\odot)}{\sqrt{g_{xx}}\sqrt{\Sigma}}, \\
   Y_1 &=& -\frac{\sqrt{g_{xx}} \, (y_0-y_\odot)}{\sqrt{g_{yy}}\sqrt{\Sigma}}, \\
   T_2 &=& \frac{\omega\sqrt{\Sigma}}{\sqrt{-g_{00}} \sqrt{-\Pi }}, \\
   X_2 &=& -\frac{(y_0-y_\odot)\sqrt{-g_{00}}}{\sqrt{\Sigma}\sqrt{-\Pi }}, \\
   Y_2 &=&  \frac{(x_0-x_\odot)\sqrt{-g_{00}}}{\sqrt{\Sigma}\sqrt{-\Pi }}, \\
   Z_3 &=& -\frac{1}{\sqrt{g_{zz}}},
\end{eqnarray}
and where we have named:
\begin{eqnarray}
   \Sigma&=&g_{xx}(y_0-y_\odot)^2+g_{yy}(x_0-x_\odot)^2\nonumber\\
   \Pi&=& g_{00}+\left[g_{xx}(y_0-y_\odot)^2+g_{yy}(x_0-x_\odot)^2\right]\omega^2.\nonumber
\end{eqnarray}

We can now solve the system (\ref{eq:3conl}) explicitly, relating the
observed quantities $\textbf{e}\bm{'_{\hat{a}}}$ to the unknowns
$\bar{\ell}^{\alpha}_{0}$ by means of the observer's comoving frame \{
$\bm{\lambda_{\hat{\alpha}}}$ \}, i.e.:
\begin{eqnarray}
\bar{\ell}_{0}^{x} &=& \frac{1}{\sqrt{g_{xx}} \sqrt{\Sigma}}
 \left[ \frac {\textbf{e}\bm{'_{\hat{1}}} (x_0-x_\odot) \sqrt{g_{yy}} 
\sqrt{-\Pi} + (y_0-y_\odot) \sqrt{g_{xx}} (\omega \sqrt{\Sigma} +
{\textbf{e}\bm{'_{\hat{2}}}} \sqrt{- g_{00}} )}
{ (\textbf{e}\bm{'_{\hat{2}}} \omega \sqrt{\Sigma} - \sqrt{-g_{00}})}
\right] 
\label{ellxo} \\
\bar{\ell}_{0}^{y}&=& \frac{1}{\sqrt{g_{yy}} \sqrt{\Sigma}} \left[ \frac {
\textbf{e}\bm{'_{\hat{1}}} (y_0-y_\odot) \sqrt{g_{xx}} \sqrt{- \Pi} -
(x_0-x_\odot) \sqrt{g_{yy}} (\omega \sqrt{\Sigma} + \textbf{e}\bm{'_{\hat{2}}}
\sqrt{- g_{00}} )}{ (\textbf{e}\bm{'_{\hat{2}}} \omega \sqrt{\Sigma} -
\sqrt{-g_{00}})}\right] \label{eq:ellyo} \\
 \bar{\ell}^{z}_{0}&=& -
\frac{\textbf{e}\bm{'_{\hat{3}}} \, \sqrt{- \Pi}} {\sqrt{g_{zz}}
\left(\textbf{e}\bm{'_{\hat{2}}} \omega \sqrt{\Sigma} - \sqrt{-g_{00}}\right).
} \label{eq:ellz0}
 \end{eqnarray}

\section{\label{sec:emtime}Identifying the emission time}
\begin{figure}
\plotone{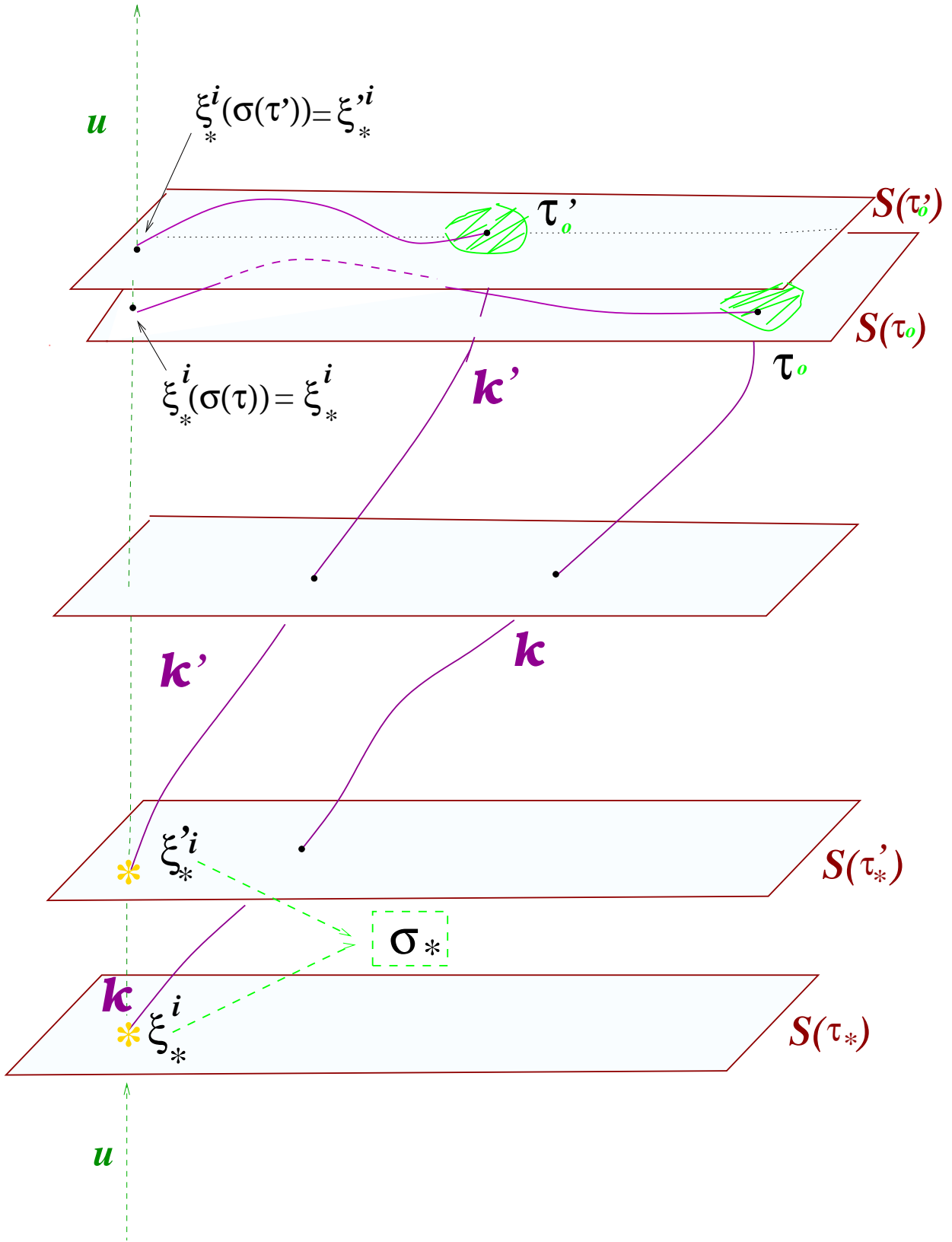}
\caption{\label{fig:emisstime}Identifying the coordinate position 
   $\xi^i_*$ of the star without proper motion and the emission 
   time $ \sigma(\tau_*)$.}
\end{figure}

Let us assume that the star has no proper motion with respect to the
given spatial coordinate grid whose origin is at the barycenter of the
Solar System.  In this case the spatial coordinates $\xi^{i}_{*}\equiv
\xi^i(\tau_*)$ of the star remain fixed with time $\tau$ and the
star's world-line is one of the curves of the congruence $C_{\bm{u}}$.
If a photon is emitted at $\tau_*$, say, then its trajectory all the
way to the observer at ($\xi^i_{(0)},\,\tau_0$) is mapped into a
spatial path on the slice $S(\tau_0)$ having the boundary conditions
fixed as explained in Section 4.  Integration along this path leads to
a solution $\xi^i=\xi^i(\sigma(\tau))$. Let the same star be observed
at a subsequent coordinate time $\tau'_0=\tau_0+\Delta\tau$; the
received photon is emitted at a coordinate time
$\tau'_{*}=\tau_*+\Delta\tau$ and its trajectory is now mapped into a
slice $S(\tau'_0)$. Evidently, this second observation implies a
different set of boundary conditions for the integration along the
second path hence the solution will be a new function
$\xi^{i'}=\xi^{i'}(\sigma(\tau))$.  Since the spatial coordinates of
the star are preserved under the mapping, they will be identified by
the value of the parameter $\sigma(\tau_*)$ such that
$\xi^{'i}(\sigma(\tau_*+\Delta\tau))=\xi^i(\sigma(\tau_*))$ (see
\figurename~\ref{fig:emisstime}).
 \clearpage
\clearpage
 
\begin{figure}
\resizebox*{0.5\textwidth}{!}{\rotatebox{-90}{\plotone{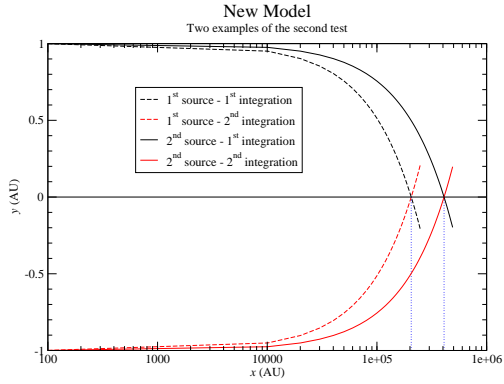}}}
\caption{\label{fig:testcross}Two examples of a test that deduces the
  distance of a star  from two observations of the same objects taken
  at opposite positions on the orbit of the observer. The first star
  is approximately at 1 parsec, the second at 2 parsecs. Each couple
  of observations gives two trajectories that cross at the expected
  distance of the objects.}
\end{figure}

A first check of the procedure is shown in
\figurename~\ref{fig:testcross}; here we assume that on the light path
acts only the gravitational field of the Sun. We considered two stars
at coordinate distances respectively of 1 and 2 parsecs.  Fixing the
boundary conditions corresponding to observations of the stars in two
symmetrically opposite directions with respect to the Sun and assuming
$\Delta\tau$ equal to six months, we find, in logarithmic units, the
points of intersections of the two integrated spatial paths.  The
actual distances are slightly less that 1 and 2 parsecs respectively
as expected since we calculate proper rather coordinate distances.
\clearpage
\clearpage

\section{\label{sec:test}Testing the model}
The perturbative model presented in the last chapter is expected to
include all terms to the order of $ (v/c)^2 $.  This means that, for a
typical velocity in the solar system of $\sim
10~\text{km}\cdot\text{s}^{-1}$, one can expect deviations from the
predictions of exact models of $\simeq2\cdot10^{-4}~\text{arcsec}$.
Equations (\ref{eq:geodint}) cannot be integrated analytically, so
numerical techniques are needed to reconstruct the light path.  We
then needed a test campaign to check the correctness and accuracy of
the model.

We used the implementation of the RADAU integrator given in
\citep{1985dcoe.conf..185E} to integrate the system of differential
equations for the null geodesic, because it has been thoroughly
tested, and can be easily implemented to reach very high orders of
accuracy.  Since the geodesic equations include both
$\bar\ell^i\equiv\dot\xi^i $ and $ \xi^i $, the system of differential
equations is of class II according to the definition used in
\citep{1985dcoe.conf..185E}, that is
$$ \textbf{y}''=F(\textbf{y}',\textbf{y},t). $$

The tests we devised verify that:
\begin{enumerate}
\item the perturbative model is self-consistent;
\item the amount of light deflection caused by each individual body of
  the Solar System, as evaluated in our perturbative model, coincides
  with that expected for an analytical Schwarzschild solution at the same
  order of accuracy and under the same observational conditions;
\item the model is able to reconstruct stellar distances.
\end{enumerate}

It is clear from these items that our tests do not consider
comparisons with other models besides the Schwarzschild solution, like
those in \citet{2002PhRvD..65f4025K} and \citet{2003AJ....125.1580K};
however, we believe that it is still too early for such
comparisons. First of all, our model is at the $(v/c)^2$ order, so
it's not strictly needed to confront it with higher order models
yet. Moreover, although with the same assumptions some results can be
recovered, the background mathematical framework is quite different
and this is another reason for not attempting a detailed
comparison. Our model is evolving to the $(v/c)^3$ order, and this
future version will be presented in a forthcoming paper and, indeed,
compared to the two works cited above.

Before illustrating each specific test, let us make some general
remarks on the numerical solution of the system of differential
equations (\ref{eq:geodint}). This system requires a set of six
boundary conditions and, as said in section~\ref{sec:incond}, the
natural choice is the three spatial coordinates of the observer at the
observation time, and the three components of the vector tangent to
the spatial line of sight.

We are dealing with a numerical problem, therefore we need to define
methods to stop the integration procedure.

One possibility is to fix a finite range of integration, the starting
point being the position of the observer and the ending point that of
the star.  This approach will be adopted for the test on stellar
distances.

The other choice is to stop the integration when the tangent vector to
the light trajectory becomes constant at the precision level of the
computer.  This method was used in the light deflection test.  In
particular, when we consider only the gravitational field generated by
the Sun, the total amount of deflection produced beyond 100~AU is
under $ 0.1~\mu$arcsec level, therefore this distance turns out to be
a pretty good choice for stopping the integration.

As a final consideration, we mention that the geodesic equations can
be adapted to various physical situations; in fact, as explained in
section~\ref{sec:metric}, the metric has not been explicited the only
requirement being that it takes the form:
\begin{equation}
g_{\alpha\beta}=\eta_{\alpha\beta}+h_{\alpha\beta}+O(h^2).\nonumber 
\end{equation}

In our approximation it is
\begin{equation}
(\eta_{\alpha\beta}+h_{\alpha\beta})\bar\ell^\alpha\bar\ell^\beta=1+O((v/c)^3),\nonumber
\end{equation}
so the first step was to verify whether the spatial vectors $ \bar\ell
$ obtained at each step of integration satisfied this unitarity
condition up to the correct order of accuracy of $ \sim 10^{-12} $.
The results show that this always happens along the integration path.

\subsection{Self-consistency test}
The basic test is whether our model satisfies spherical symmetry in
the case that the Sun is the only source of gravity.

Therefore, we took the $(v/c)^2$ approximation of the Schwarzschild
metric; the fundamental property of this space-time solution is its
spherical symmetry.  This means that the amount of light deflection
measured by an observer in a given position must depend only on the
angular distance of the celestial object from the gravitational
source.

\clearpage
\begin{figure}
\plotone{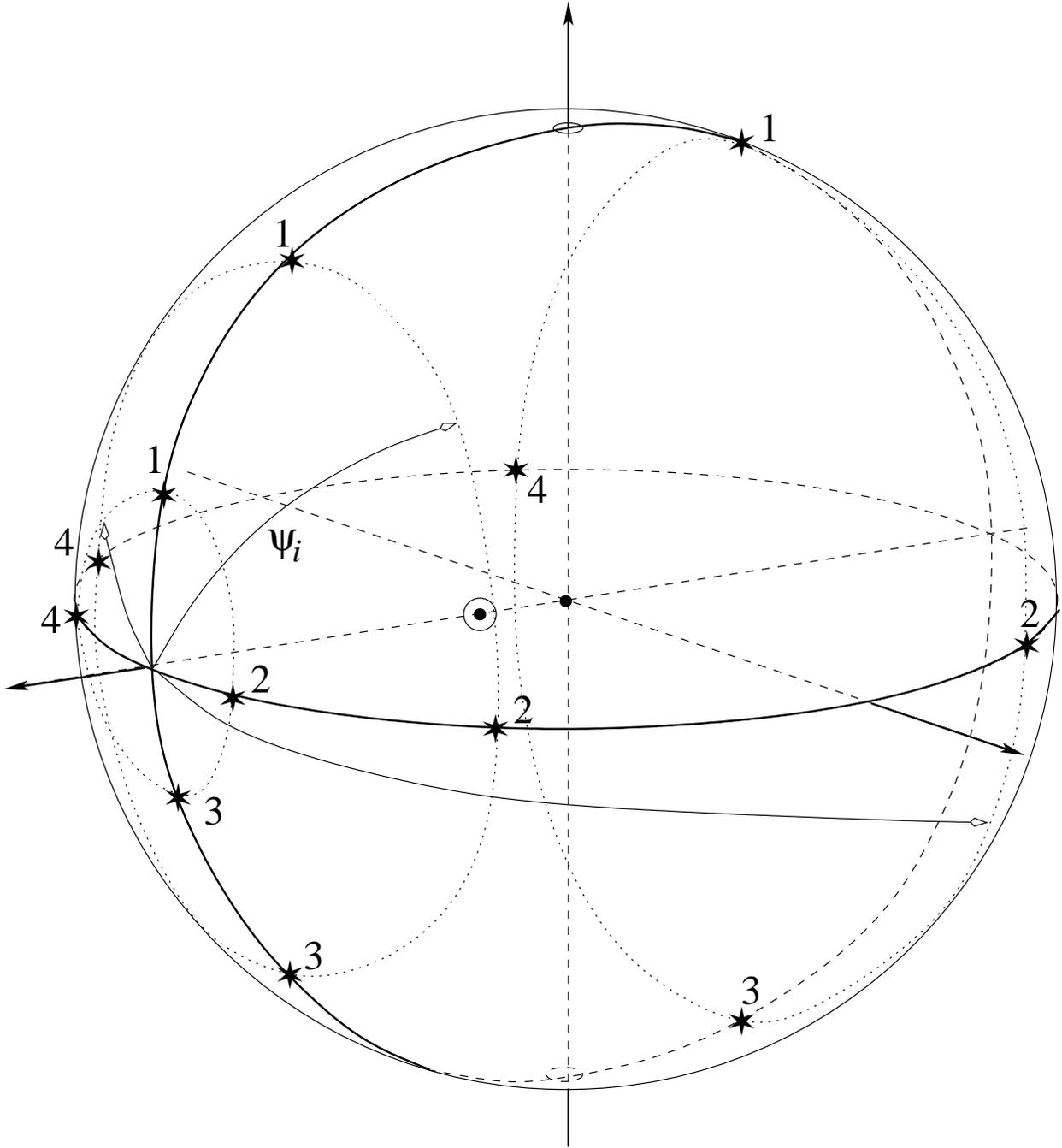}
\caption{\label{fig:self-cons}Geometry taken for the self-consistency 
  test.  For each $ \psi $ there are four stars at symmetric positions 
  with respect to the Sun-observer direction.  The angle $ \psi $ goes 
  from $ \sim 16''$ (i.e.\ limb-grazing rays) to nearly $180^\circ$. The
  observer is located at the origin of the reference frame.}
\end{figure}
\clearpage

We considered an observer at $ r_0=1 $~AU and a set of stars placed at
different angular distances $ \psi $ from the Sun.  For each $ \psi $,
we have taken four stars symmetrically positioned with respect to the
Sun as in \figurename~\ref{fig:self-cons}.

The results are reported in \tablename~\ref{tab:self-cons} where $
\delta\psi_i $ are the deflections calculated in the four cases.  They
were calculated to the $0.1$~$\mu$arcsec level and, as expected,
are the same for a given angular distance: this is a verification of
the physical consistency of the model.

The way we calculated the light deflections is discussed below.
\clearpage
\begin{deluxetable}{cc}
  \tablecolumns{2}
  \tablewidth{0pc}
  \tablecaption{Results for the self-consistency tests.  $\psi$ is the
    angular distance from the Sun, $\delta\psi_1 \ldots \delta\psi_4$
    the deflection obtained in four symmetric cases at the same $\psi$
    (only one value is reported here for each $\psi$ since they are
    exactly the same for each $\delta\psi_i$.\label{tab:self-cons}}
  \tablehead{
    \colhead{$\psi$} & \colhead{$\delta\psi_1\,\ldots\,\delta\psi_4$}
  }
\startdata
$\pz\pz0^\circ   16'\pz 5''\!\!.1428100$           & $1''\!\!.7406216$ \\
$\pz\pz1^\circ\pz 0'\pz 0''\!\!.0\phantom{000000}$ & $0''\!\!.4666385$ \\
$\pz\pz2^\circ\pz 0'\pz 0''\!\!.0\phantom{000000}$ & $0''\!\!.2333012$ \\
$\pz\pz5^\circ\pz 0'\pz 0''\!\!.0\phantom{000000}$ & $0''\!\!.0932707$ \\
  $\pz10^\circ\pz 0'\pz 0''\!\!.0\phantom{000000}$ & $0''\!\!.0465464$ \\
  $\pz45^\circ\pz 0'\pz 0''\!\!.0\phantom{000000}$ & $0''\!\!.0098314$ \\
  $\pz60^\circ\pz 0'\pz 0''\!\!.0\phantom{000000}$ & $0''\!\!.0070534$ \\
  $\pz75^\circ\pz 0'\pz 0''\!\!.0\phantom{000000}$ & $0''\!\!.0053071$ \\
  $\pz85^\circ\pz 0'\pz 0''\!\!.0\phantom{000000}$ & $0''\!\!.0044441$ \\
  $\pz95^\circ\pz 0'\pz 0''\!\!.0\phantom{000000}$ & $0''\!\!.0037316$ \\
    $105^\circ\pz 0'\pz 0''\!\!.0\phantom{000000}$ & $0''\!\!.0031248$ \\
    $120^\circ\pz 0'\pz 0''\!\!.0\phantom{000000}$ & $0''\!\!.0023511$ \\
    $135^\circ\pz 0'\pz 0''\!\!.0\phantom{000000}$ & $0''\!\!.0016868$ \\
    $170^\circ\pz 0'\pz 0''\!\!.0\phantom{000000}$ & $0''\!\!.0003563$ \\
    $175^\circ\pz 0'\pz 0''\!\!.0\phantom{000000}$ & $0''\!\!.0001778$ \\
    $179^\circ   43'   54''\!\!.8571900$           & $0''\!\!.0000095$
\enddata 
\end{deluxetable} 
\clearpage
\subsubsection{Numerical calculation of the light deflection}
\begin{figure}
\plotone{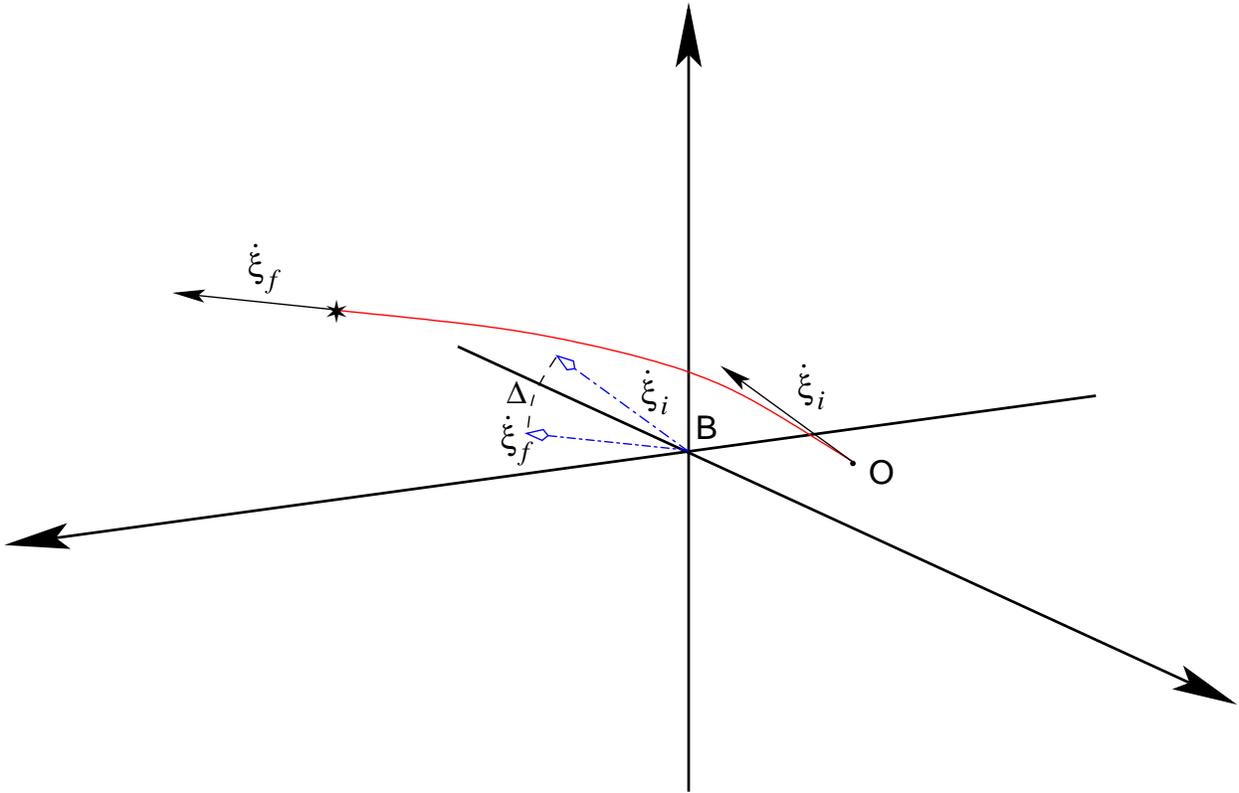}
\caption{\label{fig:defl-path}The deflection angle $ \Delta $ is the 
  angle between the initial viewing direction $\mathbf{\dot\xi}_i$ 
  and the final tangent unit vector at the end of the integration to 
  the light path $\mathbf{\dot\xi}_f$ but parallely propagated along 
  the light trajectory to the observation point.}
\end{figure}

The integrator takes as boundary conditions the Cartesian coordinates
of the observer at the time of observation and the component of the
unit vector representing the direction of the line of sight (that is
the tangent unit vector to the light path at the moment of
observation).  The quantities $\mathbf{\xi}$ and $\dot{\mathbf{\xi}}$
are recalculated at each step of integration, so the total light
deflection angle is the angle between the viewing direction
$\bar\ell^i_0$ at the observation time and the direction
$\bar\ell^i_f$ of the unit vector tangent to the light path at the end
of integration (\figurename~\ref{fig:defl-path}).

\clearpage
\clearpage

\begin{figure}
\plotone{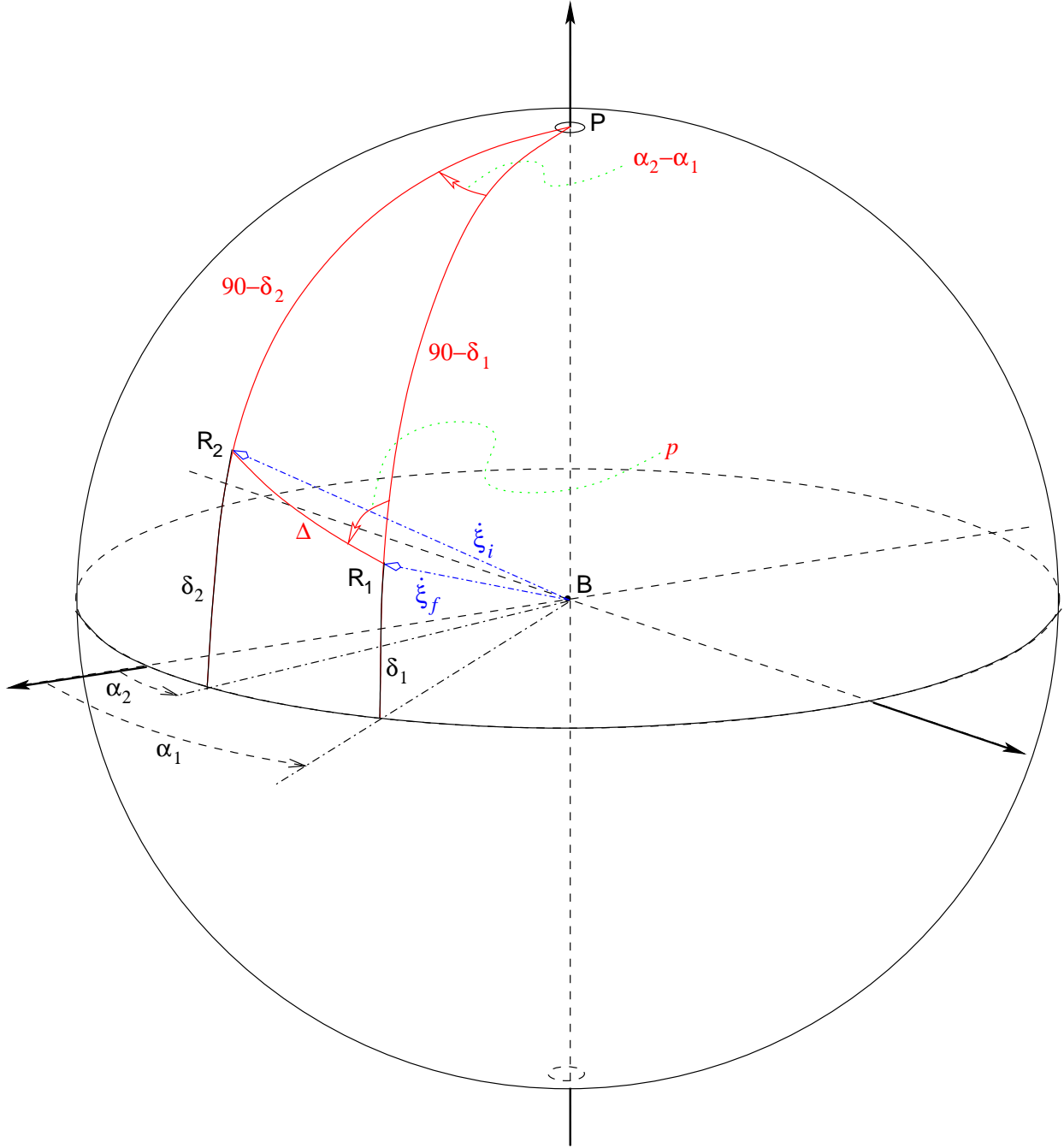}
\caption{\label{fig:ang-dist}Angular distance $\Delta$ between two 
  points $\mathsf{R}_1=(\alpha_1,\delta_1)$ and
  $\mathsf{R}_2=(\alpha_2,\delta_2)$. Here \textsf{P} is the pole 
  of the celestial sphere and $ p $ is the position angle of
  $\mathsf{R}_2$ with respect to $\mathsf{R}_1$.}
\end{figure}

The most straightforward formula for the computation of an angle
between two direction would be just
\begin{equation}
\Delta=\arccos(\mathbf{v}|\mathbf{w}),
\label{eq:delta-scalprod}
\end{equation}
where $\mathbf{v}$ and $\mathbf{w}$ are any two given unit tangent
vectors.  Unfortunately, from a numerical point of view, this formula
is of little utility, because of the smallness of the deflection
angle. In fact the numerical accuracy that we need for $ \Delta $ is
$\delta\Delta\sim1~\mu$arcsec, but for angles as small as those
expected for the relativistic deflections, namely
$\Delta\lesssim2~$arcsec, the accuracy of the built-in $\arccos$
function drops at the $10^{-4}~$arcsec level. Let us show this with a
simple test.  Consider two points $\mathsf{R}_1$ and $\mathsf{R}_2$ on
the unit celestial sphere with polar coordinates $(\alpha_1,\delta_1)$
and $(\alpha_2,\delta_2)$, and let $\alpha_1=\delta_1=\delta_2=0$ and
$\alpha_2$ variable.  Then the angle $\Delta $ between the two
directions $\textbf{v}$ and $\textbf{w}$ pointing to them is simply
given by $ \alpha_2 $.  However, application of formula
(\ref{eq:delta-scalprod}) leads to the results with poor accuracy
shown in \tablename~\ref{tab:angle-accuracy}.
\clearpage
\begin{deluxetable}{ccc}
  \tablecolumns{3}
  \tablewidth{0pc}
  \tablecaption{Test for the accuracy of the arccos function for small
    angles. The angular coordinates of two points are given as
    follows: the first one has $\alpha_1=\delta_1=0$, the second one
    has $\delta_2=0$ while $\alpha_2$ decreases. The angle $\Delta$
    between the two directions $\mathbf{v}$ and $\mathbf{w}$ pointing
    to $\mathsf{R}_1$ and $\mathsf{R}_2$ is exactly given by $
    \alpha_2 $, but is also calculated taking the arccos of the scalar
    product of the two unit vectors. The results show that when
    $\Delta\sim10^{-1}~$arcsec the accuracy of the built-in arccos is
    less than $1~\mu$arcsec.\label{tab:angle-accuracy}}
  \tablehead{
    \colhead{$\psi$} & \colhead{$\delta\psi_1\,\ldots\,\delta\psi_4$}
  }
  \tablehead{
    \colhead{$\Delta=\alpha_2~('')$} & \colhead{$\arccos~('')$} & \colhead{$\Delta-\arccos~(\mu\text{arcsec})$}
  }
\startdata
         3240.003043 &          3240.003043 & \pz\pz  0.000 \\
      \pz 324.000304 &       \pz 324.000304 & \pz    -0.001 \\
   \pz\pz  32.400030 &    \pz\pz  32.400031 & \pz    -0.069 \\
\pz\pz\pz   3.240003 & \pz\pz\pz   3.240004 & \pz    -0.485 \\
\pz\pz\pz   0.324000 & \pz\pz\pz   0.323998 & \pz\pz  2.721 \\
\pz\pz\pz   0.032400 & \pz\pz\pz   0.032382 & \pz    17.771 \\
\pz\pz\pz   0.003240 & \pz\pz\pz   0.003074 &       166.415 \\
\pz\pz\pz   0.000324 & \pz\pz\pz   0.000000 &       324.000
\enddata
\end{deluxetable}
\clearpage
A solution to this problem is to consider an approximate formula for
$\Delta$ which is better suitable for our case. From spherical
trigonometry the exact formulas for the angular distance $\Delta$ and
the position angle $p$ of two points on the unit celestial sphere
(\figurename~\ref{fig:ang-dist}) $\mathsf{R}_1=(\alpha_1,\delta_1)$
and $\mathsf{R}_2=(\alpha_2,\delta_2)$ are\footnote{We want to stress
  here that these formulas require the polar coordinates of a point
  given by the transformations $ \alpha=\arctan(x/y) $, $
  \delta=\arcsin(z/r) $, and $ r=\sqrt{x^2+y^2+z^2} $, where $ x,y,z $
  are the respective Cartesian components.}
\begin{eqnarray}
\sin\!\Delta\sin\! p\!\! &=& \!\!\cos\!\delta_2\sin(\alpha_2-\alpha_1)
\label{eq:sindsinp} \\
\sin\!\Delta\cos\! p\!\! &=& \!\!\sin\!\delta_2\cos\!\delta_1\!-\!\cos\!\delta_2\sin\!\delta_1\cos(\alpha_2\!-\!\alpha_1).
\label{eq:sindcosp}
\end{eqnarray}

\clearpage
\clearpage

The Taylor series of the sine and cosine functions about $x=0$ are
\begin{eqnarray}
\sin x &=& x-\frac{x^3}{6}+\frac{x^5}{120}+\mathcal{O}(x^7)
\label{eq:taylor-sin}\\
\cos x &=& 1-\frac{x^2}{2}+\frac{x^4}{24}+\mathcal{O}(x^6).
\label{eq:taylor-cos}
\end{eqnarray}
For $x\simeq2~\text{arcsec}\simeq10^{-5}~\text{rad}$,
$x^2\simeq10^{-10}~\text{rad}\sim10~\mu\text{arcsec}$ and
$x^3\simeq10^{-15}~\text{rad}\sim2\cdot10^{-4}~\mu\text{arcsec}$ we
can write $\forall~x\lesssim2~\text{arcsec}$
\begin{eqnarray}
\sin x &=& x+\mathcal{O}(10^{-3}\mu\text{arcsec})\label{eq:taylor-sin-muas}\\
\cos x &=& 1-\frac{x^2}{2}+\mathcal{O}(10^{-8}\mu\text{arcsec}))
\label{eq:taylor-cos-muas}.
\end{eqnarray}

Let's now write $\alpha_2-\alpha_1=\zeta\lesssim10^{-5}~\text{rad}$
and $\delta_2-\delta_1=\epsilon\lesssim10^{-5}~\text{rad}$. The
right-hand side of the (\ref{eq:sindcosp}) can be re-written, to the
correct accuracy, as
\begin{eqnarray}
 & & \sin\delta_2\cos\delta_1-\cos\delta_2\sin\delta_1\cos(\alpha_2-\alpha_1)
 \nonumber\\
 &\simeq& \sin\delta_2\cos\delta_1-
          \cos\delta_2\sin\delta_1\left(1-\frac{\zeta^2}{2}\right)
 \nonumber\\
 &=& \sin(\delta_2-\delta_1)+\frac{1}{2}\zeta^2\cos\delta_2\sin\delta_1
 \nonumber\\
 &\simeq& (\delta_2-\delta_1)+\frac{1}{2}\cos\delta_2\sin\delta_1
 \label{eq:sindcospapprox}
\end{eqnarray}

The system of equations (\ref{eq:sindsinp}) and (\ref{eq:sindcosp})
can then be approximated as
\begin{eqnarray}
\Delta\,\sin p &=& (\alpha_2-\alpha_1)\cos\delta_2
\label{eq:sdsp} \\
\Delta\,\cos p &=& (\delta_2-\delta_1)+\frac{1}{2}\cos\delta_2\sin\delta_1
\label{eq:sdcp}
\end{eqnarray}
so
\begin{eqnarray}
\Delta^2 &=& (\alpha_2-\alpha_1)^2\cos^2\delta_2+(\delta_2-\delta_1)^2+
\nonumber\\
        &{}& \frac{1}{2}\zeta^2(\delta_2-\delta_1)\cos\delta_2\sin\delta_1+
             \frac{1}{4}\zeta^4\cos^2\delta_2\sin^2\delta_1
\nonumber\\
         &=& (\alpha_2-\alpha_1)^2\cos^2\delta_2+(\delta_2-\delta_1)^2+
             \mathcal{O}(\zeta^3).
\label{eq:Delta2}
\end{eqnarray}
\clearpage
This is the formula that we shall use to compute the angle between any
two vectors\footnote{To the $\zeta^2$ order, we can neglect the
  difference between $\cos\delta_1$ and $\cos\delta_2$, in fact
  \protect\begin{eqnarray*}
    \cos\delta_2 &=& \cos(\delta_1+\epsilon) \\
    &\simeq& \cos\delta_1\left(1+\frac{1}{2}\epsilon^2\right)-
    \epsilon\sin\delta_1; \protect\end{eqnarray*} substituting this
  expression into the Eq.(\ref{eq:Delta2}) and remembering that
  $(\alpha_2-\alpha_1)=\zeta\sim\epsilon$, it is easy to show that
$$
(\alpha_2-\alpha_1)^2\cos^2\delta_2=(\alpha_2-\alpha_1)^2\cos^2\delta_1+
                                    \mathcal{O}(\epsilon^{3}).
$$
}.
\clearpage
\begin{deluxetable}{rrlccc}
  \tablecolumns{6}
  \tablewidth{0pc}
  \tablecaption{Results for the light deflection test.  The deflection
    $ \Delta_m $ is computed with a formula for the Schwarzschild
    metric for increasing values of the angular displacement $ \psi $
    of the light source from the Sun. This value is compared with that
    deduced from our model, ($ \Delta $). The difference $
    \Delta-\Delta_m $ goes rapidly under the $ 0.1~\mu $arcsec level
    for $ \psi>5^\circ $, and the magnitude of its maximum value is
    well within that expected for the $ (v/c)^2 $
    approximation.\label{tab:light-defl}}
  \tablehead{
    \multicolumn{3}{c}{$ \psi $ } & \colhead{$ \Delta $} & \colhead{$ \Delta_m $} & \colhead{$ \Delta-\Delta_m (\mu$arcsec$)$}
  }
\startdata
  $0^\circ$ & $16'$ & $\pz 5''\!\!.1428100$ & $1''\!\!.7406216$ & $1''\!\!.7406073$ & $   14.3$ \\
  $1^\circ$ & $ 0'$ & $\pz 0''\!\!.0      $ & $0''\!\!.4666385$ & $0''\!\!.4666375$ & $\pz 1.0$ \\
  $2^\circ$ & $ 0'$ & $\pz 0''\!\!.0      $ & $0''\!\!.2333012$ & $0''\!\!.2333010$ & $\pz 0.2$ \\
  $5^\circ$ & $ 0'$ & $\pz 0''\!\!.0      $ & $0''\!\!.0932707$ & $0''\!\!.0932706$ & $\pz 0.1$ \\
 $10^\circ$ & $ 0'$ & $\pz 0''\!\!.0      $ & $0''\!\!.0465464$ & $0''\!\!.0465464$ & $\pz 0.0$ \\
 $45^\circ$ & $ 0'$ & $\pz 0''\!\!.0      $ & $0''\!\!.0098314$ & $0''\!\!.0098314$ & $\pz 0.0$ \\
 $60^\circ$ & $ 0'$ & $\pz 0''\!\!.0      $ & $0''\!\!.0070534$ & $0''\!\!.0070534$ & $\pz 0.0$ \\
 $75^\circ$ & $ 0'$ & $\pz 0''\!\!.0      $ & $0''\!\!.0053071$ & $0''\!\!.0053071$ & $\pz 0.0$ \\
 $85^\circ$ & $ 0'$ & $\pz 0''\!\!.0      $ & $0''\!\!.0044441$ & $0''\!\!.0044441$ & $\pz 0.0$ \\
 $95^\circ$ & $ 0'$ & $\pz 0''\!\!.0      $ & $0''\!\!.0037316$ & $0''\!\!.0037316$ & $\pz 0.0$ \\
$105^\circ$ & $ 0'$ & $\pz 0''\!\!.0      $ & $0''\!\!.0031248$ & $0''\!\!.0031248$ & $\pz 0.0$ \\
$120^\circ$ & $ 0'$ & $\pz 0''\!\!.0      $ & $0''\!\!.0023511$ & $0''\!\!.0023511$ & $\pz 0.0$ \\
$135^\circ$ & $ 0'$ & $\pz 0''\!\!.0      $ & $0''\!\!.0016868$ & $0''\!\!.0016868$ & $\pz 0.0$ \\
$170^\circ$ & $ 0'$ & $\pz 0''\!\!.0      $ & $0''\!\!.0003563$ & $0''\!\!.0003563$ & $\pz 0.0$ \\
$175^\circ$ & $ 0'$ & $\pz 0''\!\!.0      $ & $0''\!\!.0001778$ & $0''\!\!.0001778$ & $\pz 0.0$ \\
$179^\circ$ & $43'$ & $   54''\!\!.8571900$ & $0''\!\!.0000095$ & $0''\!\!.0000095$ & $\pz 0.0$
\enddata
\end{deluxetable}
\clearpage
\subsection{The light deflection test}
A formula for the light deflection is given in
\citet{1973grav.book.....M}, for the case of the Schwarzschild metric,
as
\begin{equation}
\Delta_m=\frac{2M_\odot}{r_{\text{o}}}\sqrt{\frac{1+\cos\psi}{1-\cos\psi}},
\label{eqn:mis-defl}
\end{equation}
where $ M_\odot $ is the solar mass in geometrized units, $ r_{\text{o}}
$ is the distance of the observer from the Sun and $ \psi $ is the
angular displacement of a star from the Sun.

Since (\ref{eqn:mis-defl}) is an analytical formula to the same order
of $ (v/c)^2 $, we expect that its predictions coincide with those of
our model, namely to $\lesssim10^{-9} $~rad.

Taking the same stars and computing the light deflection using the
methods seen previously, the tests show that the difference between
the two predictions is always much less than the limit of the
approximation.  The maximum value ($ \simeq 15~\mu $arcsec) is reached
for limb grazing rays and the difference becomes rapidly less than $
0.1~\mu $arcsec for $ \psi>5^\circ $ (see
\tablename~\ref{tab:light-defl}).

\clearpage
\begin{figure}
\plotone{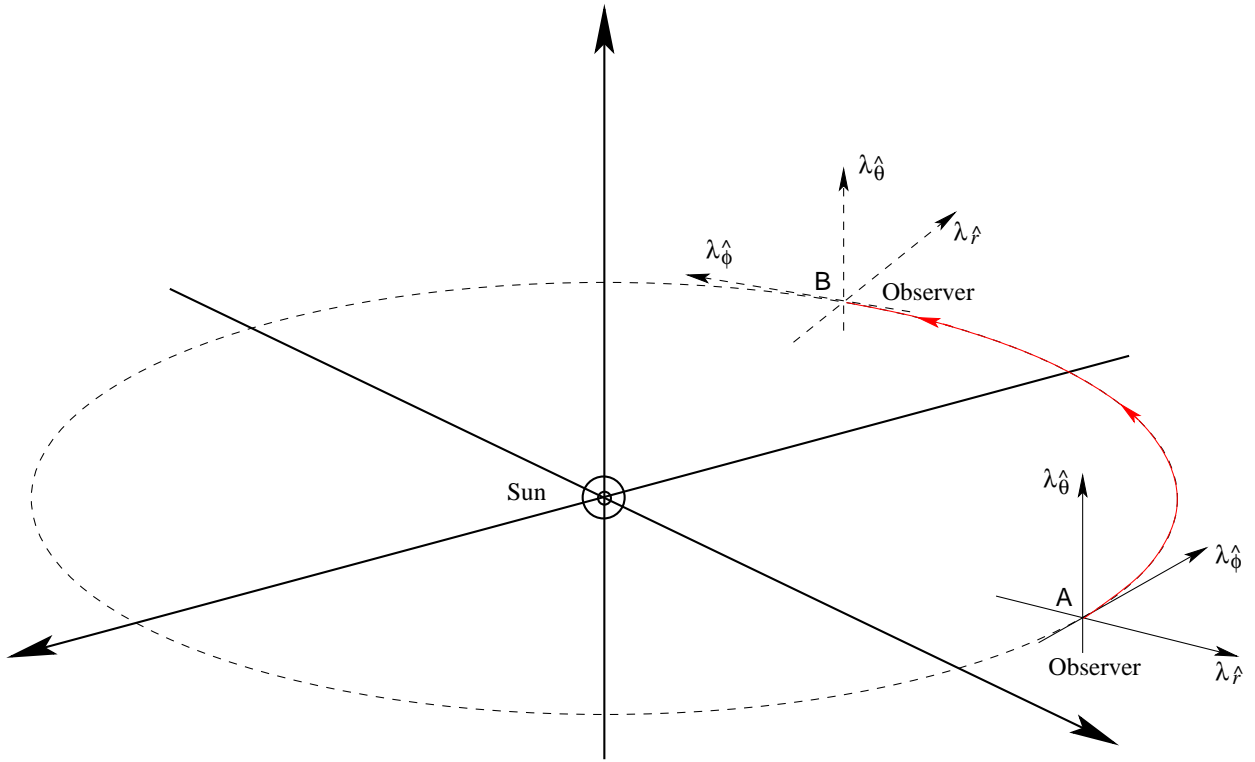}
\caption{\label{fig:phase-locked-tetrad}The spatial axes of a
  phase-locked tetrad are such that one is constantly lying in the
  Sun-observer direction, one is always tangent to the circular path
  of the observer around the Sun and the third axes is orthogonal to
  the orbital plane. The figure shows how these axes change when the
  observer moves from {\sf A} to {\sf B} along the fraction of the
  orbit represented with solid line.}
\end{figure}
\clearpage

\begin{deluxetable}{rcccc} 
  \tablecolumns{8}
  \tablewidth{0pc}
  \tablecaption{Difference (in $\mu$arcsec) between the parallax $
    p^*_e $ calculated from the exact Schwarzschild model and that
    reconstructed using our $ (v/c)^2 $ model ($ p^*_a $).  The first
    column reports the exact distance in parsec, the remaining four
    give $ (p^*_e-p^*_a) $ as obtained for each of the corresponding $
    \varepsilon_{r_c} $, the level of accuracy for the impact
    parameter $ r_c $ expressed in meters.\label{tab:comp-schw}}
  \tablehead{
    \colhead{$r^*_e~($pc$)$}    &  \multicolumn{4}{c}{$(p^*_e-p^*_a)~(\mu$arcsec$)$} \\
    \cline{2-5}
    \colhead{}    & \colhead{$\varepsilon_{r_c}=4.8$} & 
    \colhead{$\varepsilon_{r_c}=9.6\cdot10^{-2}$} & 
    \colhead{$\varepsilon_{r_c}=1.9\cdot10^{-3}$} & 
    \colhead{$\varepsilon_{r_c}=4\cdot10^{-5}$}
  }
\startdata
     $1$ & $         -262.2$ & $-12.5$ & $-17.8$ & $-17.8$ \\
    $10$ & $\pminus   124.6$ & $-19.1$ & $-17.8$ & $-17.8$ \\
   $100$ & $\pminus   150.9$ & $-23.1$ & $-17.8$ & $-17.8$ \\
  $1000$ & $\pminus   229.1$ & $-20.6$ & $-17.7$ & $-17.8$ \\
 $10000$ & $\pminus\pz 85.6$ & $-12.8$ & $-17.8$ & $-17.8$
\enddata
\end{deluxetable}
\clearpage

\subsection{\label{sec:stell-dist}Stellar distances test: comparison with a Schwarzschild model}
Recently we developed a Schwarzschild model for astrometric
observations \citep{1996arap.mthe.....V,1998AAp...332.1133D}. In that
model we express the components of the vector $ k^i $ tangent to a
null geodesic relative to the spatial axes $ \hat r $, $ \hat\theta $
and $ \hat\phi $ of a \textit{phase-locked} tetrad adapted to an
observer on a circular orbit around the Sun
(\figurename~\ref{fig:phase-locked-tetrad}), namely \citep{1992GReGr..24.1091D}
\begin{eqnarray}
\textbf{e}_{\hat r}\equiv\cos\Theta_{(\hat r,k)} &=& \pm\frac{\sqrt{\left(1-\frac{2M_\odot}{r_{\text{o}}}\right)^{-1}-\frac{\Lambda^2}{r_{\text{o}}^2}}}
                                     {\left(1-\omega\lambda\right)\sqrt{1-\frac{3M_\odot}{r_{\text{o}}}}} \label{eq:thetark} \\
\textbf{e}_{\hat\theta}\equiv\cos\Theta_{(\hat\theta,k)} &=& \pm\frac{\frac{1}{r_{\text{o}}}\sqrt{\Lambda^2-\lambda^2}}
                                          {\left(1-\omega\lambda\right)\sqrt{1-\frac{3M_\odot}{r_{\text{o}}}}} \label{eq:thetathetak} \\
\textbf{e}_{\hat\phi}\equiv\cos\Theta_{(\hat\phi,k)} &=& 
\frac{\sqrt{1-\frac{2M_\odot}{r_{\text{o}}}}\left[\lambda-\frac{\omega
r_{\text{o}}^2}{1-\frac{2M}{r_{\text{o}}}}\right]}
     {r_{\text{o}}\left(1-\omega\lambda\right)}. \label{eq:thetaphik}
\end{eqnarray}
Here $ r_{\text{o}} $ is the distance of the observer from the Sun, $
\omega $ is the coordinate angular velocity of the observer, while $
\Lambda $ and $ \lambda $ are two constants of motion of the null
geodesic\footnote{We have denoted the constants of motion
  consistently with the notation of the cited work, hence $ \lambda $
  should not be confused with the affine parameter of the null
  geodesic mentioned in section~\ref{sec:light-traj}.}.  The constants
of motion $ \lambda $ and $ \Lambda $ can be expressed as functions of
the impact parameter of the light geodesic with respect to the Sun ($
r_c $) and of the angular Schwarzschild coordinates of the star $
(\theta,\phi) $.  Finally, $ r_c $ can be implicitly expressed as a
function of the complete set of Schwarzschild coordinates of the star
$ (r,\theta,\phi) $.  This means that, given the position of a star
and of an observer (in Schwarzschild coordinates), we are able to
derive the Cartesian components of the tangent to the null geodesic at
the position of the observer by means of an almost completely
analytical procedure\footnote{The only numerical part of the
  algorithm is the calculation of $ r_c $.}.  These components allow
us to define the observables according to Eq.(\ref{eq:Chata}).

In the present model, we can reproduce the same physical situation and
the same observables by taking $ h_{\alpha\beta} $ as the
approximation to the Schwarzschild metric and by adopting the tetrad,
defined by Eqs.(\ref{eq:lamb0c})--(\ref{eq:lamb3c}), of an observer on
a circular orbit. Hence we can use those observables and the position
of the observer as the boundary conditions needed to integrate
backwards the set of differential equations~(\ref{eq:geodint}).  It
was this complementarity of the two algorithms that was exploited for
our test.

We start by giving the ``true'' position of a star $
(r^*,\theta^*,\phi^*) $ and two symmetric observers (with respect to
the Sun) $ (r_1,\theta_1,\phi_1) $ and $ (r_2,\theta_2,\phi_2) $ in
the ``Schwarzschild model'', from which we calculate the tetraedal
components of the local light directions for both observers using
Eqs.(\ref{eq:thetark})--(\ref{eq:thetaphik}).  Let's call those
components $
(\textbf{e}_{r_1},\textbf{e}_{\theta_1},\textbf{e}_{\phi_1}) $ and $
(\textbf{e}_{r_2},\textbf{e}_{\theta_2},\textbf{e}_{\phi_2}) $.  Note
that $ (r_1,\theta_1,\phi_1) $ and $
(\textbf{e}_{r_1},\textbf{e}_{\theta_1},\textbf{e}_{\phi_1}) $
represent the boundary conditions needed in our model to integrate the
light trajectory for the first observer, and analogously $
(r_2,\theta_2,\phi_2) $ and $
(\textbf{e}_{r_2},\textbf{e}_{\theta_2},\textbf{e}_{\phi_2}) $ are the
boundary conditions for the second observer.

The point of intersection of the two paths obtained in this way should
be the position of the star at $ (r^*,\theta^*,\phi^*) $; however we
expect to find slightly different coordinates $
(r^*_a,\theta^*_a,\phi^*_a) $.  Indeed, there are two reasons for this
difference: the approximation of our perturbative model which is up to
$ (v/c)^2 $, and possible numerical errors.  The latter should always
be maintained below the level of the intrinsic approximation.

In this test we have taken the two observers at opposite positions on
the orbit of the Earth around the Sun (i.e. $ r_1=r_2=r_{\text{o}} $,
$ \theta_1=\theta_2=\pi/2 $ and $ \phi_2=\phi_1+\pi $).  The stars are
also on the orbital plane ($ \theta^*=\pi/2 $) in such a way the two
observers are symmetrically placed with respect to the Sun-star
direction ($ \phi^*=(\phi_1+\phi_2)/2 $).  Their distances ($r^*$)
range from $ 1 $~pc to $ 10 $~kpc.

With this configuration it is easy to calculate the parallax of a star
as $ p^*=1/r^* $~rad (the distances being in AU), and so it is equally
easy to determine the numerical accuracy of the position in terms of
angles by simply subtracting the parallax $ p^*_e $ of the exact model
from the approximate one, $ p^*_a $, reconstructed from the numerical
integrations described above.

The results in \tablename~\ref{tab:comp-schw}, show that there is an
accuracy floor at about $ 18~\mu $arcsec, as one could expect given
the intrinsic level of approximation of the model.  This table also
shows that an accuracy on the determination of $ r_c $ to $
\sim10^{-3} $~m at least, is needed to reach the theoretical floor for
$ (p^*_e-p^*_a) $. We stress that in our case, where $ r_c\simeq
r_0\simeq 1.5\cdot 10^{11}~$m, this means that the numerical accuracy
needed is $ \varepsilon_{r_c}/r_c\sim10^{-14} $, pushing to the edge
the typical accuracy for a standard double precision number; however
to this level the numerical contribution to the error is contained to
the $ 0.1~\mu $arcsec level.

\clearpage
\begin{figure}
\plotone{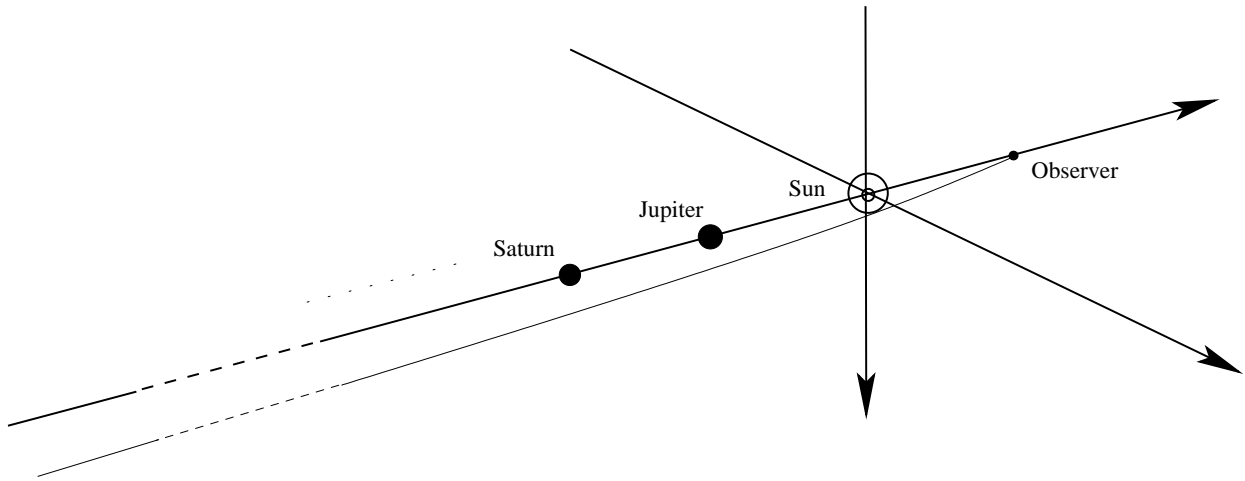}
\caption{\label{fig:n-bod-defl}The geometrical configuration of the 
  deflection test with more than one body is such that all the planets
  are aligned behind the Sun and along the line joining the Sun with
  the observer.}
\end{figure}
\clearpage

\begin{figure*}
\includegraphics[scale=0.65]{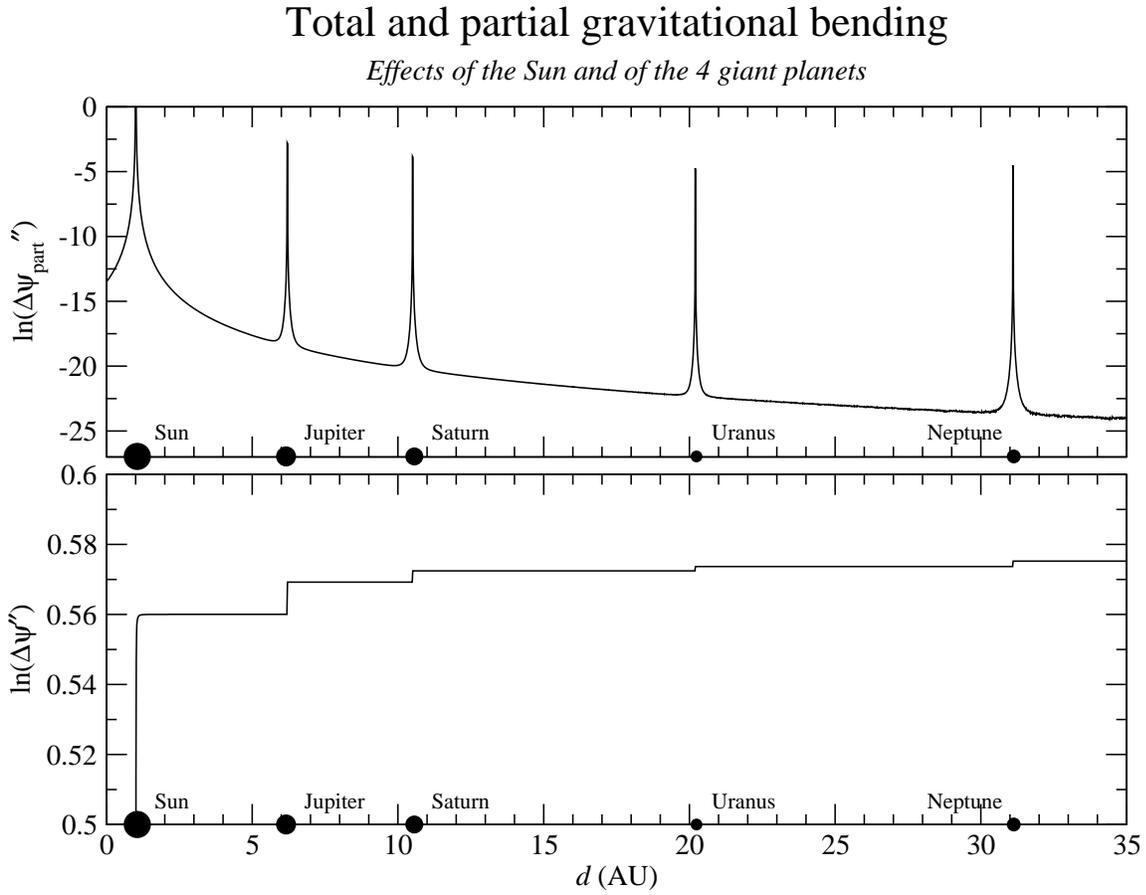}
\caption{\label{fig:graph-defl}The upper plot shows the natural
  logarithm of the partial deflection $\Delta\psi_{\text{part}}$
  (expressed in arcseconds) of the light ray due to the Sun and the
  four major planets of the Solar System, while the other gives the
  total deflection. Both are plotted against the approximate distance
  $d$ from the observer in Astronomical Units. All the bodies
  considered are drawn as circles in their approximate locations along
  the $ x $-axis.}
\end{figure*}
\clearpage

\section{\label{sec:n-bod-test}N-body tests}
The tests described in section~\ref{sec:test} consider the Sun as the
unique source of gravity. This is because in that section we wanted to
compare our $ (v/c)^2 $ model with an exact one (or at least with a
model which was already tested at the level of accuracy we require) and
the obvious choice was the Schwarzschild model.

However, our model has been built to include an arbitrary number of
gravity sources.  We shall then subject our model with more than one
gravitating body to similar tests to see whether in this case the
results are consistent with expectations. This will also serve as the
proper testing ground for our forthcoming $ (v/c)^3 $ model and
whatever others would be available.  Indeed, to the best of our
knowledge, we do not know of any extensive numerical testing for
multi-body relativistic models.

\subsection{Deflection due to a variable number of aligned bodies}
\label{sec:n-body-defl}
In this test we have calculated the total deflection due to the Sun
plus a variable number of planets. The geometry is
depicted in \figurename~\ref{fig:n-bod-defl}; here the planets are
aligned behind the Sun with respect to the observer and the light ray
was assumed to be grazing the solar limb. The results are reported in
\tablename~\ref{tab:n-bod-defl} and in
\figurename~\ref{fig:graph-defl}.

The first row of the table provides the total deflection in four
different cases: the first one gives the result for the Sun plus
Jupiter, the second one for the Sun, Jupiter and Saturn, and so on.

\clearpage
\begin{deluxetable}{lcccc}
  \tablecolumns{5}
  \tablewidth{0pc}
  \tablecaption{Results for the multi-body deflection test. All the
    numbers are expressed in arcseconds. The first row gives the total
    deflection due to the gravitational pull of the Sun, plus that of
    the planet indicated in the corresponding column heading, and plus
    that of all the other planets on its left.  The value of the
    deflection in the last row, $\Delta\psi_{\astrosun}$, is for a
    light ray grazing the limb of the Sun, i.e.\ where the distance of
    closest approach to the center of the Sun is
    $r_{\mathrm{c}}=R_{\astrosun}=696\,000$~km.\label{tab:n-bod-defl}}
  \tablehead{
    \colhead{} & \colhead{Jupiter (\jupiter)} & \colhead{Saturn (\saturn)} & 
    \colhead{Uranus (\uranus)} & \colhead{Neptune (\neptune)}
  }
\startdata
$ \Delta\psi_i $ & $ 1.7509921 $ & $ 1.7510398 $ & $ 1.7510436 $ & $ 1.7510465 $ \\
$ \Delta\equiv\Delta\psi_{\astrosun}-\Delta\psi_i $ & $ 0.0002699 $ & $ 0.0003176 $ & $ 0.0003214 $ & $ 0.0003243 $ \\
$ \Delta_\mathrm{schw} $ & $ 0.0002695 $ & $ 0.0003172 $ & $ 0.0003210 $ & $ 0.0003239 $ \\
$ \Delta-\Delta_\mathrm{schw} $ & $ 0.0000004 $ & $ 0.0000004 $ & $ 0.0000004 $ & $ 0.0000004 $ \\
\cutinhead{$ \Delta\psi_{\astrosun}=1.7507222 $}
\enddata
\end{deluxetable}
\clearpage

The second row shows the differences between the deflection due to the
Sun (i.e.\ the value $\Delta\psi_{\astrosun}$) and the amount of total
deflection immediately above. This difference gives the deflection due
to the planets alone.

The same effect can be approximately estimated calculating the light
deflection due to each planet, with the analytical formula for the
Schwarzschild solution, and summing them up.  This is what is
represented in the third row.  Finally the fourth row simply reports
the differences among the values in second row and those in third.

We stress here that the predictions contained in the third row are not
well justified from a theoretical point of view.  However they can be
considered {\it reasonably\/} close to reality; hence this and the
following tests just tell us that our model behaves ``well'' with
respect to a reasonable (but not exact) model under the same physical
situation.

\clearpage
\begin{figure}
\plotone{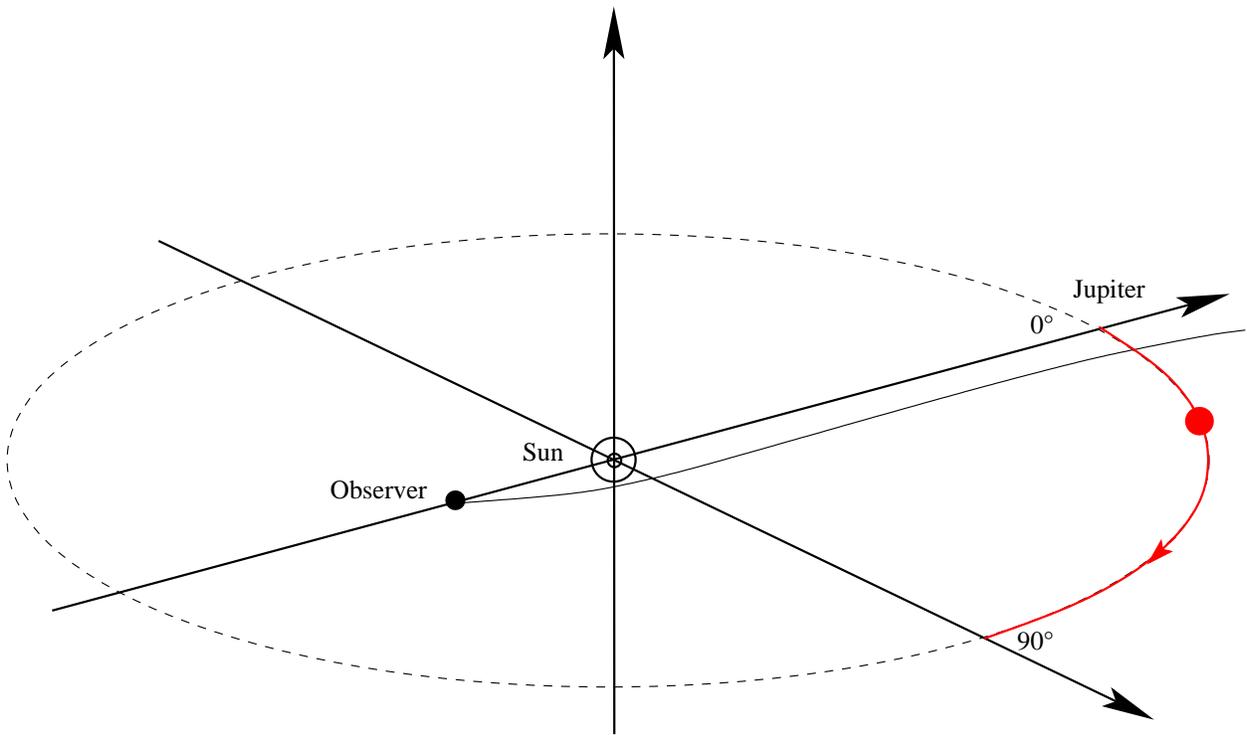}
\caption{\label{fig:n-bod-defl-vp}
  The total deflection due to the Sun and Jupiter was calculated
  varying the position of Jupiter along its orbit.}
\end{figure}

\clearpage
\begin{figure}
\plotone{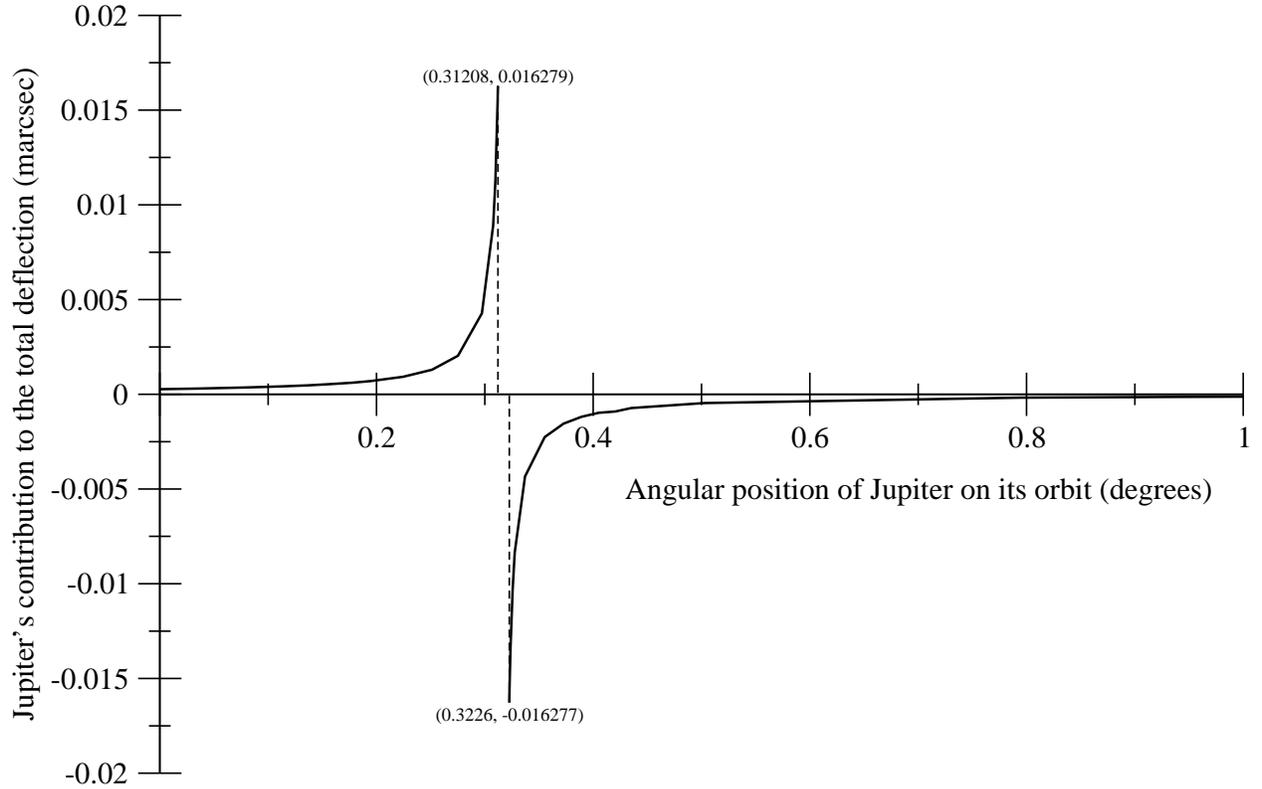}
\caption{\label{fig:defl-jup-vp}
  When Jupiter is aligned with the Sun and the observer its position
  on the orbit is $ \alpha=0^\circ $. Its contribution to the total
  deflection increases approximately to $ 16.28 $~mas for $
  \alpha\simeq0^\circ\!\!.31208 $, where the light path grazes both
  the Sun and the planet leaving them on the same side. For $
  \alpha\simeq0^\circ\!\!.3226 $ the light ray still grazes both of
  them, but passing in the middle, so that Jupiter's contribution
  subtracts to the total deflection. With Jupiter in the range $
  0^\circ\!\!.31208<\alpha<0^\circ\!\!.3226 $ the photons are stopped
  by the planet.}
\end{figure}
\clearpage

\begin{deluxetable}{ccc}
  \tablecolumns{3}
  \tablewidth{0pc}
  \tablecaption{Results for the test of
    section~\ref{sec:defl-two-bod}. The angular positions of Jupiter
    are in degrees and the deflections in
    arcseconds.\label{tab:defl-two-bod}}
  \tablehead{
    \colhead{Angle} & \colhead{$\Delta\psi_\mathrm{tot}$} & \colhead{$\Delta\psi_{\jupiter}$}
  }
\startdata
 0.0000000 & 1.7509921 &  0.0002699 \\
 0.0355333 & 1.7510261 &  0.0003039 \\
 0.0794550 & 1.7510823 &  0.0003601 \\
 0.1123663 & 1.7511401 &  0.0004179 \\
 0.1376200 & 1.7511988 &  0.0004766 \\
 0.1589100 & 1.7512629 &  0.0005407 \\
 0.1776667 & 1.7513355 &  0.0006133 \\
 0.1946242 & 1.7514202 &  0.0006980 \\
 0.2247327 & 1.7516472 &  0.0009250 \\
 0.2512588 & 1.7520185 &  0.0012963 \\
 0.2752403 & 1.7527570 &  0.0020348 \\
 0.2972935 & 1.7549956 &  0.0042734 \\
 0.3077281 & 1.7596352 &  0.0089130 \\
 0.3097728 & 1.7620437 &  0.0113215 \\
 0.3118041 & 1.7661978 &  0.0154756 \\
 0.3120065 & 1.7667852 &  0.0160630 \\
 0.3120672 & 1.7669702 &  0.0162480 \\
 0.3120773 & 1.7670014 &  0.0162792 \\
 0.3226012 & 1.7344451 & -0.0162771 \\
 0.3226110 & 1.7344754 & -0.0162468 \\
 0.3227480 & 1.7348868 & -0.0158354 \\
 0.3229435 & 1.7354393 & -0.0152829 \\
 0.3237245 & 1.7373081 & -0.0134141 \\
 0.3256688 & 1.7404386 & -0.0102836 \\
 0.3276016 & 1.7423752 & -0.0083470 \\
 0.3370993 & 1.7463871 & -0.0043351 \\
 0.3553339 & 1.7484676 & -0.0022546 \\
 0.3726774 & 1.7491742 & -0.0015480 \\
 0.3892489 & 1.7495309 & -0.0011913 \\
 0.4051432 & 1.7497466 & -0.0009756 \\
 0.4204370 & 1.7498130 & -0.0009092 \\
 0.4351937 & 1.7499953 & -0.0007269 \\
 0.5025188 & 1.7502596 & -0.0004626 \\
 0.7945559 & 1.7505427 & -0.0001795 \\
 0.9401342 & 1.7505846 & -0.0001376 \\
 1.1236807 & 1.7506159 & -0.0001063 \\
 1.3762332 & 1.7506413 & -0.0000809 \\
 1.5891499 & 1.7506548 & -0.0000674 \\
 1.9463344 & 1.7506696 & -0.0000526 \\
 2.2474695 & 1.7506778 & -0.0000444 \\
 2.5127875 & 1.7506832 & -0.0000390 \\
 3.5539031 & 1.7506957 & -0.0000265 \\
 7.9518754 & 1.7507110 & -0.0000112 \\
11.2547126 & 1.7507143 & -0.0000079 \\
15.9423686 & 1.7507167 & -0.0000055 \\
19.5572137 & 1.7507177 & -0.0000045 \\
30.0733476 & 1.7507193 & -0.0000029 \\
39.7151372 & 1.7507200 & -0.0000022 \\
54.7655820 & 1.7507206 & -0.0000016 \\
67.3801351 & 1.7507209 & -0.0000013 \\
78.9125108 & 1.7507211 & -0.0000011 \\
90.0000000 & 1.7507213 & -0.0000009
\enddata
\end{deluxetable}
\clearpage
\begin{deluxetable}{ccc}
  \tablecolumns{3}
  \tablewidth{0pc}
  \tablecaption{The distance of maximum approach of the light ray to
    Jupiter ($ r_\mathrm{min} $) is calculated in the two extremal
    cases in which the photons pass nearby the limb of the planet. It
    is always slightly more than the planet's actual radius $
    R_{\jupiter}=71398 $~km
    \citep{1999ssd.book......M}.\label{tab:max-appr}}
  \tablehead{
    \colhead{Row} & \colhead{$ r_\mathrm{min} $} & \colhead{$ \Delta\psi_{\jupiter} $}
  }
\startdata
18 & 71461~km & $ \phantom{-}0''\!\!.0162792 $ \\
19 & 71474~km & $ -0''\!\!.0162771 $ \\
\enddata
\end{deluxetable}
\clearpage
\subsection{\label{sec:defl-two-bod}Deflection of two bodies with variable relative position}
In this test we considered only two bodies, the Sun and Jupiter, and
calculated the total light deflection experienced by a light ray
grazing the solar limb and positioning Jupiter in several different
places along its orbit. In particular, the planet's position varied in
a range of $ 90 $~degrees, from conjunction to quadrature
(\figurename~\ref{fig:n-bod-defl-vp}).

The results (\tablename~\ref{tab:defl-two-bod} and
\figurename~\ref{fig:defl-jup-vp}) show that the contribution of
Jupiter, as it should be, adds to that of the Sun when the two masses
are both ``on the same side'' with respect to the light path; their
their influences subtract when the light passes between the bodies.
Moreover, when the light ray grazes the Jovian limb (rows number 18
and 19 of \tablename~\ref{tab:defl-two-bod}), the contribution to the
total deflection due to the planet approaches the predicted
theoretical value of $ 16.28 $~marcsec
\citep{2000tmcs.conf..314D,2000tmcs.conf..265K} as shown in
\tablename~\ref{tab:max-appr}.

Finally, it is also worth noting that the amount of Jupiter's
contribution to the total deflection is of the order of $ 1~\mu
$arcsec when the planet is $ 90 $~degrees far from the light path,
again in very good agreement with the theoretical predictions
\citep{2000tmcs.conf..265K}.
\clearpage
\begin{deluxetable}{lccccc}
  \tabletypesize{\scriptsize}
  \tablecolumns{6}
  \tablecaption{\label{tab:n-bod-par}}
  \tablehead{
    \colhead{} & \colhead{\astrosun} & \colhead{\astrosun+\jupiter} & 
    \colhead{\astrosun+\jupiter+\saturn} & \colhead{\astrosun+\jupiter+\saturn+\uranus} & 
    \colhead{\astrosun+\jupiter+\saturn+\uranus+\neptune}
  }
\startdata
\cutinhead{distance $\sim 1$~pc}
$d~($AU$)$                             & 206261.338 & 206259.698 & 206259.216 & 206259.142 & 206259.054 \\*
$p_i~('')$                             & 1.00001775 & 1.00002571 & 1.00002804 & 1.00002840 & 1.00002883 \\*
$p_i-p_{\astrosun}~(\mu$arcsec$)$      & \nodata    & -7.95 & -10.29 & -10.65 & -11.07 \\*
$\Delta\psi_i~(\mu$arcsec$)$           & \nodata    & -8.00 & -10.30 & -10.60 & -11.10 \\
\cutinhead{distance $\sim 10$~pc}
$d~($AU$)$                             & 2062283.55 & 2062120.14 & 2062072.27 & 2062064.94 & 2062056.3 \\*
$p_i~('')$                             & 0.10001777 & 0.10002569 & 0.10002802 & 0.10002837 & 0.10002879\\*
$p_i-p_{\astrosun}~(\mu$arcsec$)$      & \nodata    & -7.93 & -10.25 & -10.60 & -11.02 \\*
$\Delta\psi_i~(\mu$arcsec$)$           & \nodata    & -8.00 & -10.30 & -10.60 & -11.00 \\
\cutinhead{distance $\sim 100$~pc}
$d~($AU$)$                             & 20589921.6 & 20573645.4 & 20568881.5 & 20568152.5 & 20567292.3 \\*
$p_i~('')$                             & 0.01001777 & 0.01002569 & 0.01002801 & 0.01002837 & 0.01002879 \\*
$p_i-p_{\astrosun}~(\mu$arcsec$)$      & \nodata    & -7.93 & -10.25 & -10.60 & -11.02 \\*
$\Delta\psi_i~(\mu$arcsec$)$           & \nodata    & -7.99 & -10.30 & -10.60 & -11.00 \\
\cutinhead{distance $\sim 1000$~pc}
$d~($AU$)$                             & 202664764  & 201098829  & 200644596  & 200575248  & 200493473  \\*
$p_i~('')$                             & 0.00101776 & 0.00102569 & 0.00102801 & 0.00102837 & 0.00102879 \\*
$p_i-p_{\astrosun}~(\mu$arcsec$)$      & \nodata    & -7.93 & -10.25 & -10.60 & -11.02 \\*
$\Delta\psi_i~(\mu$arcsec$)$           & \nodata    & -7.99 & -10.30 & -10.60 & -11.00 \\
\cutinhead{distance $\sim 10000$~pc}
$d~($AU$)$                             & 1751489760 & 1641052370 & 1611285280 & 1609051510 & 1603803810 \\*
$p_i~('')$                             & 0.00011777 & 0.00012569 & 0.00012801 & 0.00012819 & 0.00012861 \\*
$p_i-p_{\astrosun}~(\mu$arcsec$)$      & \nodata    & -7.93 & -10.25 & -10.42 & -10.84 \\*
$\Delta\psi_i~(\mu$arcsec$)$           & \nodata    & -7.99 & -10.30 & -10.60 & -11.00 \\
\enddata
\end{deluxetable}
\clearpage
\subsection{Stellar distances revisited: effect of many bodies}
As a final test, we have repeated the experiment of
section~\ref{sec:stell-dist}, but with an increasing number of aligned
planets as in section~\ref{sec:n-body-defl} (the sequence of planets
was kept the same).  However, unlike the previous case, in this test
we cannot compare our model with any other, so this is in fact another
test of self-consistency for the $ (v/c)^2 $ model. The sense of this
statement will be clarified below.

\clearpage
\begin{figure}
\begin{center}
\epsscale{.85}
\plotone{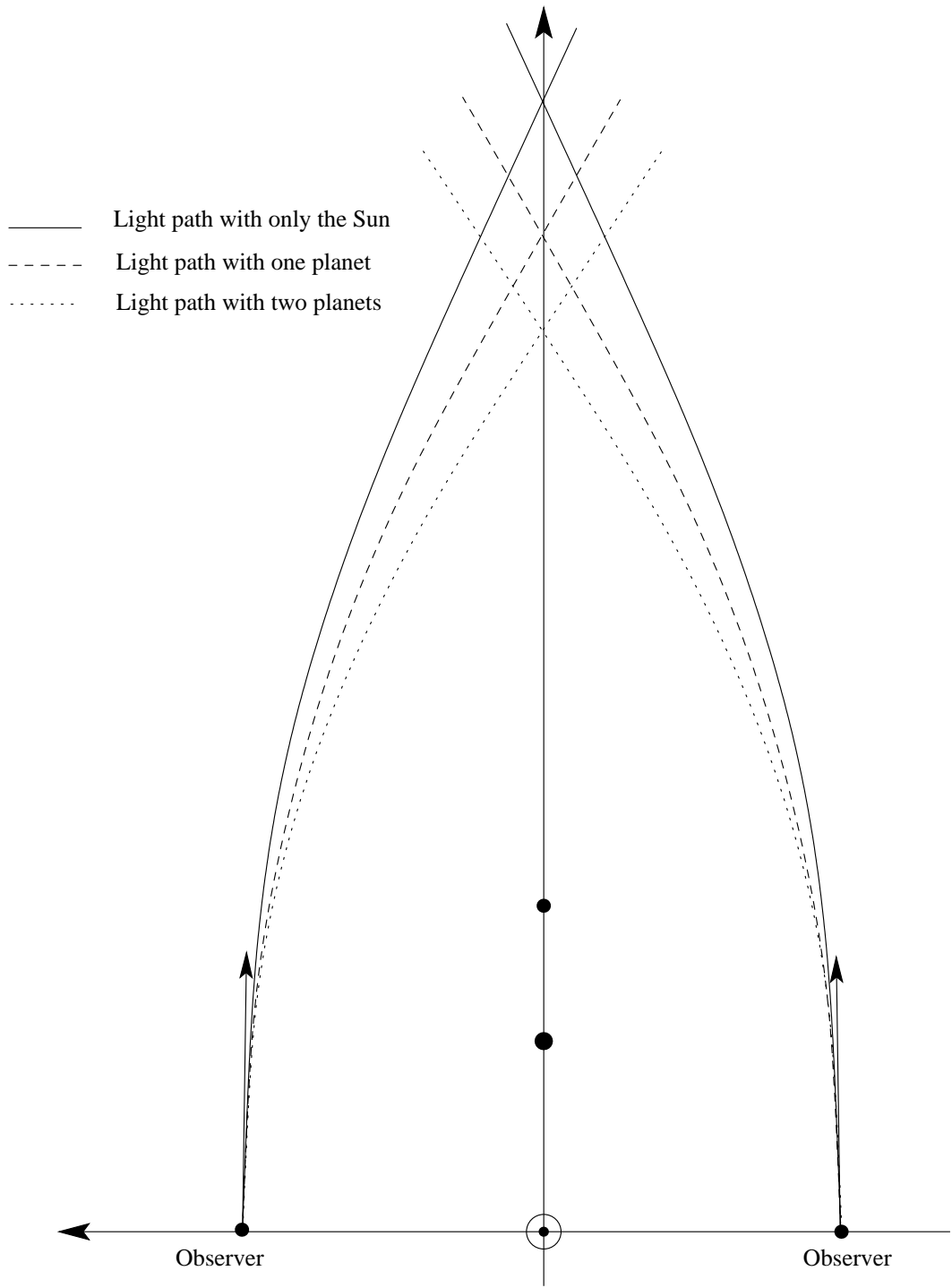}
\end{center}
\caption{\label{fig:n-bod-par}The same initial conditions (i.e.\ viewing directions) should
  result in a decreased distance of the star when other planets are
  added.}
\end{figure}
\clearpage

We consider the same boundary conditions as those used for the test
with the Schwarzschild model. These were unchanged as we added Jupiter,
Saturn and the other planets.

Therefore, the expectation was the the distance to a star decreased
as we added more planets (and the corresponding parallax increased)
because of the increased total deflection
(\figurename~\ref{fig:n-bod-par}).

The results are reported in \tablename~\ref{tab:n-bod-par}. Stellar
distances, as before, ranged from 1~pc to 10~kpc. The first row in
each set of 4 is the derived distance, $ d $, in AU (i.e.\ that
obtained by the intersection of the two geodesics), the second is the
corresponding parallax $ p=206265/d $~arcsec\footnote{Since, as in
  section~\ref{sec:n-body-defl}, each column gives the results for a
  given number of planets (i.e., the single Sun in first column, the
  Sun+Jupiter in the second, and so on), we will refer to the
  quantities in the first row as $ d_{\astrosun} $, $ d_{\jupiter} $,
  \ldots and to those in the second row as $ p_{\astrosun} $, $
  p_{\jupiter} $, \ldots}.  The third row contains the difference
between $ p_{\astrosun} $ and the parallax of the corresponding
column. For instance, in the case of Jupiter, the value is $
p_{\astrosun}-p_{\jupiter}\equiv\Delta p_{\jupiter} $. Finally, the
fourth row provides the contribution $ \Delta\psi_i $, ($
i=\jupiter,\saturn,\uranus,\neptune $) of the planets to the total
deflection.  This contribution is calculated independently, following
the method used in section~\ref{sec:n-body-defl}.

The self-consistency of the model assures that these last two values
coincide up to numerical errors.

The results show that, as we add more planets, the distances decrease
confirming our qualitative expectations. Moreover, the numerical
residuals are $ \Delta p_i-\Delta\psi_i\lesssim 10^{-1}~\mu $arcsec ($
i=\jupiter,\saturn,\uranus,\neptune $), and this again is compatible
with our accuracy for double precision numbers.

\section{Conclusions}
We have developed a general relativistic astrometric n-body model
capable of following a light ray all the way from the emitting star to
an observer (satellite) orbiting around the Sun.

This model has an intrinsic accuracy of $\sim0.1$~milliarcsecond
($(v/c)^2$) and its computer implementation was validated via a
thorough test campaign that proved: {\it a)\/} the self-consistency of
the model; {\it b)\/} that the amount of light deflection, in the case
of a spherical, non-rotating, single-body metric, coincides with that
produced by an analytical solution; {\it c)\/} that, under the same
assumptions, our model is able to reconstruct stellar distances; {\it
  d)\/} that in the case of a multi-body configuration, for which we
don't have any exact or even well-founded analytical approximation,
the outcomes for the light deflection and the reconstruction of
stellar distances are consistent with the results from
semi-quantitative derivations.

Finally, this model will be the natural test-bed for the more advanced
astrometric model, accurate to $(v/c)^3$, which will be presented in a
forthcoming article.

\begin{acknowledgments}
Work partially supported by the Italian Space Agency (ASI) under
contracts ASI I/R/117/01, and by the Italian Ministry for Research
(MIUR) through the COFIN 2001 program.
\end{acknowledgments}

\appendix
\section{\label{app:mapping}Mathematical description of the mapping procedure}
Here we describe rigorously the mathematical procedure used to obtain
the spatial projection of the light ray on the slice $S(\tau_0)$.  The null geodesic crosses
each slice $S(\tau)$ at a point with coordinates
$\xi^i=\xi^i(\lambda(\tau))$; but this point also belongs to the
unique normal to the slice $S(\tau)$, crossing it with a value of the
parameter $\sigma=\sigma(\xi^i(\lambda),\,\tau)\equiv
\sigma_{\xi^i(\lambda)}(\tau)$. (\figurename~\ref{fig:sigmalambda}).

Let us now define the one-parameter local diffeomorphism:
\begin{equation}
\phi_{\Delta\sigma}\equiv\phi_{(\sigma_{\xi^i(\lambda)}(\tau_0)- \sigma_{\xi^i(\lambda)}(\tau))}\,:\,\mit\Upsilon\cap S(\tau) \to S(\tau_0)\label{eq:diff}
\end{equation}
which maps each point of the null geodesic $\mit\Upsilon$ to the point
on the slice $S(\tau_0)$ which one gets to by moving along the unique
normal through the point $\mit\Upsilon(\lambda)\cap S(\tau)$, by a
parameter distance
$\Delta\sigma=\sigma_{\xi^i(\lambda)}(\tau_0)-\sigma_{\xi^i
  (\lambda)}(\tau)$ (\figurename~\ref{fig:sigmalambda}). Since the
spatial coordinates $\xi^i$ are Lie-transported along the normals to
the slices, then the points in $S(\tau_0)$, which are images of those
on the null geodesic under $\phi_{\Delta\sigma}$, have coordinates
$(\phi_{\Delta\sigma}(\mit\Upsilon(\lambda)\cap S(\tau)))^i=\xi^i$.
The curve in $S(\tau_0)$ which is the image of $\mit\Upsilon$ under
$\phi_{\Delta\sigma}$, is:

\begin{equation}
\bar{\mit\Upsilon}\equiv\phi_{\Delta\sigma}\circ\mit\Upsilon ;
\label{eq:curveim}
\end{equation}
it has tangent vector:
\begin{equation}
\dot{\bar{\mit\Upsilon}}^\alpha=\dot{\Big(\phi_{\Delta\sigma}^*\circ
k\Big)}^\alpha= \frac{\partial\xi^\alpha(\sigma(\tau_0))}{\partial\xi^\beta
(\sigma(\tau))}k^\beta\equiv\ell^\alpha.\label{eq:tangveim}
 \end{equation}
 \clearpage
\begin{figure}
\plotone{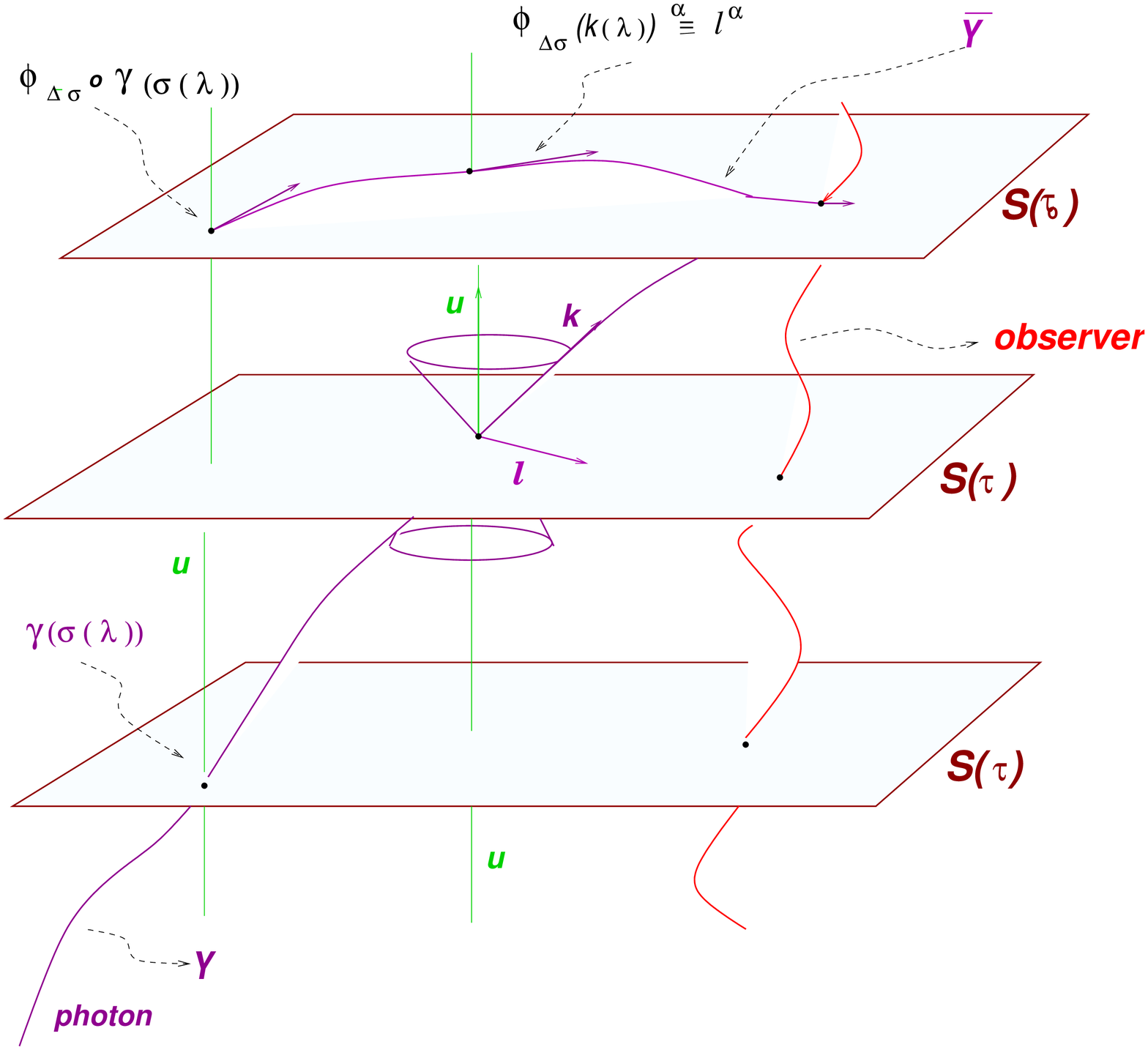}
\caption{\label{fig:fig5}The curve $\bar{\mit\Upsilon}$ will be the image of
  the null geodesic $\mit\Upsilon_k$ under the diffeomorphism
  $\phi_{\Delta \sigma}$.}
\end{figure}
\clearpage
The map (\ref{eq:diff}) acts on a 4-dimensional manifold with images
in a 3-dimensional one hence the coordinates of the target points
($\mit\Upsilon\cap S(\tau)$) and those of the image points on
$S(\tau_0)$ are respectively:
\begin{eqnarray}
\xi^\alpha(\sigma(\tau))&=&\tau\delta^\alpha_0+\xi^i(\sigma(\tau))\delta^\alpha_i\nonumber\\
\xi^\alpha(\sigma(\tau_0))&=&\xi^\alpha(\sigma(\tau))-\tau\delta^\alpha_0.\nonumber
\end{eqnarray}
From this and (\ref{eq:curveim}) it follows that (\figurename~\ref{fig:fig5}). 
\begin{equation}
\ell^\alpha=\tilde\gamma^\alpha{}_\beta k^\beta\,, \label{eq:ell}
\end{equation}
hence the curve $\bar{\mit\Upsilon}$ is the spatial projection of the
null geodesic on the slice $S(\tau_0)$ at the time of observation.

\section{\label{app:tetr-comp}Explicit expression for the tetrad components}
The following expressions represent the coordinate components of each
tetrad vector $\lambda_{\hat{a}}$ (a= 1, 2, 3) with respect the
coordinate basis $\bm{\partial_\alpha}$ ($\alpha=0, x, y, z$) when the
observer moves on the orbit which has general barycentric component
$X_s$, $Y_s$, $Z_s$ (to be specified):
\begin{eqnarray*}
   T_s &=& 1 \\
   T_1 &=& 0 \\
   X_1 &=& \mp\frac{Y_{s}\sqrt{g_{yy}}}{\sqrt{g_{xx}}\sqrt{g_{xx}X_{s}^{2}+g_{yy}Y_{s}^{2}}} \\
   Y_1 &=& \pm\frac{X_s\sqrt{g_{xx}}}{\sqrt{g_{yy}}\sqrt{g_{xx}X_{s}^{2}+g_{yy}Y_{s}^{2}}} \\
   Z_1 &=& 0 \\
\end{eqnarray*}
\begin{eqnarray*}
   T_2 &=& \mp\frac{\sqrt{(g_{xx}X_{s}^{2}+g_{yy}Y_{s}^{2})/(g_{00}+g_{zz}Z_{s}^{2})}}
                   {\sqrt{g_{00}+g_{xx}X_{s}^{2}+g_{yy}Y_{s}^{2}+g_{zz}Z_{s}^2}} \\
   X_2 &=& \pm\frac{X_s\sqrt{(g_{00}+g_{zz}Z_{s}^{2})/(g_{xx}X_{s}^{2}+g_{yy}Y_{s}^{2})}}
                   {\sqrt{g_{00}+g_{xx}X_{s}^{2}+g_{yy}Y_{s}^{2}+g_{zz}Z_{s}^2}} \\
   Y_2 &=& \pm\frac{Y_s\sqrt{(g_{00}+g_{zz}Z_{s}^{2})/(g_{xx}X_{s}^{2}+g_{yy}Y_{s}^{2})}}
                   {\sqrt{g_{00}+g_{xx}X_{s}^{2}+g_{yy}Y_{s}^{2}+g_{zz}Z_{s}^2}} \\
   Z_2 &=& \mp\frac{Z_s\sqrt{(g_{xx}X_{s}^{2}+g_{yy}Y_{s}^{2})/(g_{00}+g_{zz}Z_{s}^{2})}}
                   {\sqrt{g_{00}+g_{xx}X_{s}^{2}+g_{yy}Y_{s}^{2}+g_{zz}Z_{s}^2}} \\
\end{eqnarray*}
\begin{eqnarray*}
   T_3 &=& \mp\frac{Z_s\sqrt{g_{zz}}}{\sqrt{g_{00}}\sqrt{g_{00}+g_{zz}Z_{s}^{2}}} \\
   X_3 &=& 0 \\
   Y_3 &=& 0 \\
   Z_3 &=& \pm\frac{\sqrt{g_{00}}}{\sqrt{g_{zz}}\sqrt{g_{00}+g_{zz}Z_{s}^{2}}}.
\end{eqnarray*}

\nocite{1980PhRvD..22.2947E,1999PhRvD..60l4002K,1986PhRvD..34.2246A,
        1990CQGra...7.1733B,1973grav.book.....M,1992PhRvD..45.1017D,
        1993PhRvD..47.3124D,1994PhRvD..49..618D,1991ercm.book.....B,
        1993PhRvD..48.1451K}



\end{document}